\documentclass{aa}

\usepackage{graphicx}
\usepackage{amsmath}
\usepackage{amssymb}
\usepackage{pjournal}
\usepackage{natbib}
\usepackage{multirow}

\bibpunct{(}{)}{;}{a}{}{,}

\begin{document}


\title{Estimating the total infrared luminosity of galaxies up to $z \sim 2$
from mid- and far-infrared observations}

\author{N.~Bavouzet \inst{1}
  \and H.~Dole \inst{1}
  \and E.~Le Floc'h \inst{2}
  \and K.I.~Caputi \inst{3}
  \and G.~Lagache \inst{1}
  \and C.S.~Kochanek \inst{4}
}

\institute{Institut d'Astrophysique Spatiale (IAS), b\^at 121,
Universit\'e Paris-Sud 11 and CNRS (UMR 8617), F-91405 Orsay Cedex, France
\and
Institute for Astronomy, University of Hawaii, 2680 Woodlawn Drive, Honolulu, HI 96822
\and
Institute of Astronomy, Swiss Federal Institute of Technology (ETH Z\"urich), CH-8093,
Z\"urich, Switzerland
\and
Department of Astronomy, Ohio State University, 140 West 18th Avenue, Columbus, OH 43210
}

\date{Received 1 January 2001 / Accepted 1 January 2001}

\abstract
{}
{We present the observed correlations between rest-frame 8, 24, 70 and 160~$\mu$m 
monochromatic luminosities and measured total infrared luminosities 
$L_\textrm{IR}$ of galaxies detected by Spitzer.}
{Our sample consists of 372 star-forming galaxies with individual detections 
and flux measurements at 8, 24, 70 and 160~$\mu$m. We have
spectroscopic redshifts for 93\% of these sources, and accurate photometric
redshifts for the remainder. We also used a stacking analysis
to measure the IR fluxes of fainter sources at higher redshifts.}
{We show that the monochromatic mid and far-infrared luminosities
are strongly correlated with the total infrared luminosity and our
stacking analysis confirms that these correlations also hold at higher
redshifts. We provide relations between monochromatic luminosities and
 total infrared luminosities $L_\textrm{IR}$ that should be reliable
up to $z\sim2$ ($z\sim1.1$) for ULIRGs (LIRGs). In particular, we can
predict $L_\textrm{IR}$ with accuracies of 37\% and 54\% from the 8 and 24
$\mu$m fluxes, while the best tracer is the 70~$\mu$m flux.
Combining bands leads to slightly more accurate estimates.
For example, combining the 8 and 24~$\mu$m luminosities predicts
$L_\textrm{IR}$ with an accuracy of 34\%. Our results are generally
compatible with previous studies, and the small changes
are probably due to differences in the sample selection criteria.
We can rule out strong evolution in dust properties with redshift up to $z\sim 1$.
Finally, we show that infrared and sub-millimeter observations
are complementary means of building complete samples of star-forming
galaxies, with the former being more sensitive
for $z \lesssim 2$ and the latter at higher $z \gtrsim 2$.}
{}


\titlerunning{Estimating the total infrared luminosity of galaxies up to $z \sim 2$}

\maketitle


\section{Introduction}

Bright infrared galaxies play an important role in understanding the
evolution of galaxies because a large fraction of the energy
from star formation is reprocessed by dust and only visible
in the infrared. This emission is dominated by the
luminous infrared galaxies (LIRGs, defined by 
$10^{11} L_\odot < L_\textrm{IR} < 10^{12} L_\odot$) at $z>0.7$ 
\citep[e.g.][]{le_floch2005} and
by ultra-luminous infrared galaxies (ULIRGs, defined by
$L_\textrm{IR} > 10^{12} L_\odot$) at $z>2$ \citep[e.g.][]{caputi2007}.
The LIRGs and ULIRGs are massive star-forming
galaxies \citep{swinbank2004,caputi2006b} but some of the emission may
also come from embedded AGN \citep[e.g.][]{alonso-herrero2006}. 
For systems dominated by star formation, the total infrared luminosity
$L_\textrm{IR}$ emitted between a few $\mu$m and 1~mm is a good tracer
of the star formation rate \citep{kennicutt98}. Thus estimating the
infrared luminosity of galaxies is important for quantifying star formation activity.

Measuring the spectral energy distribution (SED)
of infrared galaxies also probes physical properties such
as their dust temperature, or grain size and composition.
Models of nearby objects
using IRAS, ISO and Spitzer have shown that the typical infrared
galaxy SED peaks between 60 and 150$\mu$m, depending on the dust
temperature \citep[e.g. the Spitzer Nearby Galaxy Survey, ][]{kennicutt2003}.

The far-infrared bands on Spitzer (the MIPS 70 and 160$\mu$m bands)
and Akari (FIS\footnote{Far-Infrared Surveyor}) are well-suited to
study this peak. Unfortunately, far-infrared detectors generally
have less sensitivity and poorer resolution (given the fixed telescope
aperture) than mid-infrared detectors, frequently leading to surveys
that are either limited by confusion \citep{dole2004,dole2004a,frayer2006}
or noisier than their mid-IR counterparts.

Ground based sub-millimeter facilities, such as e.g. the SHARC-2 instrument
 at the CSO\footnote{Caltech Submillimeter Observatory},
SCUBA\footnote{Submillimeter Common-User Bolometer Array}
 mounted on the JCMT\footnote{James Clerk
Maxwell Telescope} or APEX--2a 
and LABOCA\footnote{Large Apex BOlometer CAmera} 
recently installed on the Atacama Pathfinder EXperiment, have made a number
of important first steps \citep[e.g.][]{chapman2005},
exploiting the fact that the peak of the infrared emission is shifted
into the sub-millimeter bands at high redshifts. However, sub-millimeter surveys 
have had difficulties in building large galaxy samples because of their
limited sensitivity, and they are biased towards detecting objects 
with colder dust temperatures than the typical infrared selected 
ULIRG \citep[e.g.][]{chapman2004,pope2006}.

\citet{chary2001} and \citet{takeuchi2005a} have shown that mid-infrared
monochromatic luminosities, by which we mean measurements through filters
of modest widths, are sufficiently well-correlated with the total infrared
luminosity to provide an estimate of $L_\textrm{IR}$. In both cases, the
relations were calibrated on local samples and then extrapolated to higher redshifts.
This extrapolation is potentially dangerous, because the dominant source populations 
are considerably different
at higher redshifts (ULIRGs) and locally (non-LIRGs).
\citet{marcillac2006} studied these correlations at
higher redshifts ($0.4<z<1.2$) with a small sample of 49 
15~$\mu$m-selected galaxies, but their estimates of $L_\textrm{IR}$ were
strongly model-dependent. \citet{symeonidis2006} studied a similar sample
of galaxies selected at 70~$\mu$m, but did not examine how the
far-infrared emission was correlated with the mid-infrared emission.

In this paper we determine the correlations between $(\nu
L_\nu)_{8\mu\textrm{m, rest}}$, $(\nu L_\nu)_{24\mu\textrm{m, rest}}$,
$(\nu L_\nu)_{70\mu\textrm{m, rest}}$, $(\nu L_\nu)_{160\mu\textrm{m,
rest}}$ and $L_\textrm{IR}$. We will calibrate the relations using a
large sample of galaxies at moderate redshifts ($z<0.5$) and then test
their validity at higher redshifts ($z<2$) using a stacking analysis.
We define our samples in Section 2 and we explain our approach to
estimating $L_\textrm{IR}$ in Section 3. The correlations and their 
implications are discussed in Sections 4 and 5.
We adopted an $H_0=71\textrm{ km s}^{-1}\textrm{Mpc}^{-1}$,
$\Omega_M=0.27$, $\Omega_\Lambda=0.73$ cosmology throughout this paper.


\section{Data sample and processing}

Our main imaging datasets come from the IRAC \citep{fazio2004} and 
MIPS \citep{rieke2004} instruments on board the {\it
Spitzer Space Telescope} \citep{werner2004}. We worked on three
different fields: the Bo\"otes field of the NOAO Deep
Wide Field Survey \citep[NDWFS;][]{jannuzi99}, the extragalactic First
Look Survey (FLS), and the {\it Chandra} Deep Field--South
(CDFS). Characteristics of the IRAC and MIPS observations in these
fields are given by \citet{eisenhardt2004,lacy2004,fazio2004}
 and \citet{papovich2004, papovich2006, dole2004,
frayer2006}, respectively. We also used, for the stacking analysis, the
ultra-deep data from the GOODS survey in the GOODS/CDFS and GOODS/HDFN
fields and the galaxy sample of \citet{caputi2007}.

\subsection{Optical spectra and photometry}

\subsubsection{Bo\"otes and FLS}
In the 8~deg$^2$ Bo\"otes field, we use redshifts from the first
season of the AGN and Galaxy Evolution Survey (AGES, Kochanek et
al. in preparation). Samples
of galaxies were restricted to $R \lesssim 20$ with well-defined
sub-samples over a broad range of wavelengths. In particular,
all galaxies with $S_{24\mu \textrm{m}} \ge 1 \textrm{ mJy}$
and $R \le 20$ were targeted. The R-band magnitude limit restricts
this sample to $z<0.5$.
\citet{papovich2006} conducted quite similar observations in the FLS field (4~deg$^2$), 
but the infrared selection is deeper ($S_{24} > 0.3 \textrm{ mJy}$).
The redshift completenesses are $\sim 90\%$ for sources with $i \le 21$ and
$S_{24} \ge 1 \textrm{ mJy}$, and 35\% for $i \le 20.5$ and
$0.3 \textrm{ mJy} \le S_{24} < 1 \textrm{ mJy}$ sources.
Both of these redshift surveys used the 300 fiber Hectospec instrument
on the MMT \citep{fabricant2005}.

After excluding stars and quasars, we have 846 (504) galaxies from
the AGES (FLS) samples with spectroscopic redshifts and 24~$\mu$m
detections. All of them are in the common field of view at every infrared wavelength
(from 3.6 through 160~$\mu$m).

\subsubsection{CDFS}
The CDFS has been observed by many instruments covering a wide
wavelength range (from X-rays to radio).
\citet{wolf2004} published a photometric redshifts catalog 
of 63501 sources brighter than $R \sim 25$ over an area of 0.5~deg$^2$ of the CDFS
(COMBO-17). However, as pointed
by \citet{le_floch2005}, only the subsample of sources with $R<24$ and $z<1.2$ can
be securely used. From this catalog, we selected objects unambiguously
identified as galaxies with $R \le 22$. The redshifts of these sources 
go up to $z=0.8$ and are accurate to $\le$ 2\%. Following \citet{wolf2004} suggestion,
we excluded from
our sample 7 surprisingly bright objects with $R \le 19$ and $0.4 \le z \le 1.1$.
The final CDFS sample is composed of 1747 optical galaxies with
accurate photometric redshift determinations. Note that this 
sample is a complete optical flux-limited sample without the
joint optical/24~$\mu$m criteria used for the Bo\"otes and FLS samples.

\begin{table*}[t!]
\caption{Galaxy Sample Completeness by Wavelength}
\label{table:detection:gal}
\begin{center}
\begin{tabular}{l c c c c c c c c c c}
\hline\hline
\multirow{2}*{Field} & Initial & Initial & \multirow{2}{*}{3.6$\mu$m} & \multirow{2}{*}{4.5$\mu$m} &
\multirow{2}{*}{5.8$\mu$m} & \multirow{2}{*}{8.0$\mu$m} & \multirow{2}{*}{24$\mu$m} & 
\multirow{2}{*}{70$\mu$m} & \multirow{2}{*}{160$\mu$m} & Final \\
  & selection & sample size & & & & & & & & sample size \\
\hline
Bo\"otes & Opt. + 24$\mu$m &  846 &  846 &  846 &  834 & 841 & 821$^1$ &  428    & 236    & 185 \\
FLS      & Opt. + 24$\mu$m &  504 &  504 &  504 &  495 & 503 & 494$^1$ &  378    & 187    & 162 \\
CDFS     & Opt.            & 1767 & 1731 & 1688 & 1055 & 707 & 438$^2$ &  54$^3$ & 38$^3$ & 25  \\
\hline
Total    &                 &      &      &      &      &     &         &         &        & 372 \\
\hline
\end{tabular}
\end{center}
\begin{scriptsize}

\hspace{0.5cm}$^1$ Extended sources are not detected with the PSF fitting photometry method and
are thus excluded from our sample.

\hspace{0.5cm}$^2$ Only sources first detected at 8$\mu$m are analyzed because of the high
number of sources in the initial sample. This avoids false cross-identifications of sources.

\hspace{0.5cm}$^3$ Only sources first detected at 24$\mu$m are analyzed because of the high
number of sources in the initial sample. This avoids false cross-identifications of sources.
\end{scriptsize}
\end{table*}

\subsection{Spitzer Photometry}
\label{photometry}

We measured infrared fluxes from 3.6 through 160 $\mu$m for all the
galaxies in the Bo\"otes, FLS and CDFS samples using the following procedures.

For each IRAC channel, we constructed a mosaic and its error maps
from the post-BCD images using the MOPEX package\footnote{{\tt http://ssc.spitzer.caltech.edu/postbcd/index.html}}.
After a fine re-centering on the IR sources, we
measure the flux in an aperture 1\farcs5 in radius. We measured
the background in a 2\arcsec{} wide annulus. We optimized the radius of
the annulus over the range from 7\arcsec{} to 27\arcsec{} in 
steps of 1\arcsec{} by finding the annulus with the minimum 
$| \gamma | \sigma $ where $\gamma$ and $\sigma$ are the
skewness and dispersion of its pixels.
After subtracting the background, we corrected
the source flux for the finite aperture size relative to the PSF by 
factors of 1.74, 1.83, 2.18 and 2.44 at 3.6, 4.5, 5.8 and 8.0 $\mu$m,
respectively. Both extraction errors and photon noise were taken into
account in the estimation of uncertainties. Sources are considered as
securely detected if their signal to noise ratio is greater that 3. 
The fluxes were calibrated using calibration factors of 36.04, 34.80, 36.52
and 37.20~$\mu$Jy/(MJy/sr) at 3.6, 4.5, 5.8 and 8.0~$\mu$m, 
and the uncertainties in these factors are less than 3\% (IRAC
Handbook\footnote{{\tt
http://ssc.spitzer.caltech.edu/IRAC/dh/}}).

The MIPS observations were done using the scan map mode and were then
reduced with the DAT \citep{gordon2005}. We measured the 24 $\mu$m fluxes
by fitting a PSF to the sources. A PSF model was build for each map
using 10 bright, isolated sources.
Our PSF fitting method is not designed to measure the flux of extended 
sources, so we excluded 35 nearby extended sources from the
Bo\"otes/FLS samples. At 70 and 160 $\mu$m, we used aperture photometry.
After a re-centering on the sources, we determined the flux 
within apertures of 18\arcsec{} and 25\arcsec{}, backgrounds 
in annuli of [50\arcsec{}--70\arcsec{}] and 
[80\arcsec{}--110\arcsec{}], and applied aperture corrections
of $1.68$ and $2.29$ for 70 and 160~$\mu$m, respectively.
We used flux calibration factors of 0.0447 MJy/sr/$U_{24}$,
702~MJy/sr/$U_{70}$, and 44.6~MJy/sr/$U_{160}$ for the 
three bands \citep{gordon2006} where the $U_x$ are the
standard units of the MIPS maps and the calibrations are
uncertain by 4, 7 and 13\%. We estimated our 3-$\sigma$
detection limits of 23, 14 and 11 mJy at 70 $\mu$m and 92, 79 and 
59 mJy at 160 $\mu$m in the Bo\"otes, FLS and CDFS fields respectively
by measuring the scatter $\sigma_r$ in the fluxes measured at random 
positions on each map after rejecting outliers to the distributions.
This procedure should account for both instrumental and confusion noise. 
A source is considered as detected if its flux exceeds $3\sigma_r$,
and Tab.~\ref{table:detection:gal} summarizes the detection rates
for each MIPS band.

It may appear strange that several sources detected at 160~$\mu$m
are not detected at 70~$\mu$m (Tab.~\ref{table:detection:gal}).
Two reasons can be invoked to explain this situation. First,
sources can be intrinsically brighter at 160~$\mu$m, and this
can compensate for the sensitivity difference. If 
$S_{70\textrm{,lim}}$ and $S_{160\textrm{,lim}}$ are the
detection limits, then a $z=0$ source detected at 160~$\mu$m but missed 
at 70~$\mu$m must satisfy
\begin{equation}
\frac{(\nu L_\nu)_{160\mu \textrm{m, rest}}}{(\nu L_\nu)_{70\mu \textrm{m, rest}}}
 > \frac{70}{160} \times \frac{S_{160\textrm{,lim}}}{S_{70\textrm{,lim}}},
\end{equation}
corresponding to minimum ratios of 1.7, 2.4, 2.4 in the Bo\"otes, FLS
and CDFS respectively. Such rest-frame 160/70 colors are not extreme
and are observed in our sample (see Fig.~\ref{figure:evol_color}).
It becomes easier to satisfy such a criterion at higher redshifts
because the observed-frame 70~$\mu$m flux will tend to be smaller than the 
rest-frame frame flux, while the observed-frame 160~$\mu$m flux
will tend to be larger than the rest-frame flux. Second,
at both 70 and 160~$\mu$m, we are detecting sources below the 
completeness level \citep{dole2004,frayer2006}, so
not all the 160~$\mu$m sources are detected at 70~$\mu$m because
they are below the completeness limit at this wavelength.

Our final sample consists of 372 galaxies that are detected at 8, 24, 70 
and 160~$\mu$m. In the CDFS, the final sample is very small compared to 
its initial size (see Tab.~\ref{table:detection:gal}) because we started
with an optically-selected sample rather than a joint optical/24$\mu$m-selected
sample as we used for the Bo\"otes and  FLS fields.

\subsection{Stacking analysis}
\label{stacking}

\begin{table*}[t!]
\caption{Results for the stacking analysis in the Bo\"otes field. The mean fluxes are 
given in mJy where ``$\cdots$'' means that no source were detected. The uncertainties 
are the jackknife uncertainties.}
\label{table:stack_bootes}
\centering
\begin{tabular}{c c c c c c c c}
\hline\hline
$S_{24\mu\textrm{m}}$ bin & \multirow{2}{*}{Redshift bin} & \multirow{2}{*}{$N_s$} & $<S_{8}>$ & $<S_{24}>$  & $<S_{70}>$ & $<S_{160}>$ \\
(mJy) & & & (mJy) & (mJy) & (mJy) & (mJy) \\
\hline
                 & $0<z<0.25$   & 191 & $0.48\pm 0.02$ & $1.16\pm 0.16$ & $16.6\pm 0.9$ & $38.0\pm 2.4$ \\
$0.8<S_{24}<1.5$ & $0.25<z<0.5$ & 113 & $0.25\pm 0.01$ & $1.18\pm 0.16$ & $14.9\pm 1.0$ & $38.6\pm 2.8$ \\
                 & $0.5<z<1$    &  17 & $0.13\pm 0.01$ & $1.18\pm 0.16$ & $12.1\pm 3.7$ & $34.3\pm 9.4$ \\
\hline
                 & $0<z<0.25$   & 148 & $0.67\pm 0.03$ & $2.07\pm 0.42$ & $23.2\pm 1.2$ & $42.4\pm 2.7$ \\
 $1.5<S_{24}<3$  & $0.25<z<0.5$ & 65  & $0.35\pm 0.02$ & $2.00\pm 0.39$ & $22.9\pm 1.7$ & $73.7\pm 28.2$ \\
                 & $0.5<z<1$    & 10  & $0.29\pm 0.08$ & $2.01\pm 0.43$ & $14.7\pm 4.7$ & $32.0\pm 9.3$ \\
\hline
                 & $0<z<0.25$   & 39 & $0.96\pm 0.06$ & $4.59\pm 1.66$ & $41.9\pm 3.7$ & $53.7\pm 5.2$ \\
  $3<S_{24}<10$  & $0.25<z<0.5$ & 11 & $0.61\pm 0.09$ & $3.90\pm 1.15$ & $25.7\pm 6.0$ & $43.1\pm 12.9$ \\
                 & $0.5<z<1$    & 3  & $0.61\pm 0.24$ & $4.91\pm 2.51$ & $14.5\pm 5.2$ &   $\cdots$    \\
\hline
\end{tabular}
\end{table*}

\begin{table*}[t!]
\caption{Results of the stacking analysis of the \citet{caputi2007} sample
in the CDFS and HDFN fields. The mean fluxes are in mJy and the uncertainties
are the jackknife uncertainties. All stacked sources are securely detected 
with a (photometric) signal-to-noise ratio greater than 3.}
\label{table:stack_karina}
\centering
\begin{tabular}{c c c c c c c}
\hline\hline
\multirow{2}{*}{Redshift bin} & \multirow{2}{*}{$z_\textrm{med}$} &
\multirow{2}{*}{$N_s$} & $S_{8}$ & $S_{24}$  & $S_{70}$ & $S_{160}$ \\
 & & & (mJy) & (mJy) & (mJy) & (mJy) \\
\hline
 $0<z<0.3$   & 0.20 &  78 & $0.297\pm 0.069$ & $0.517\pm 0.116$ & $9.32\pm 2.56$ & $8.74\pm 1.99$ \\
 $0.3<z<0.6$ & 0.46 & 193 & $0.058\pm 0.006$ & $0.205\pm 0.013$ & $2.41\pm 0.48$ & $9.14\pm 1.89$ \\
 $0.6<z<0.9$ & 0.73 & 283 & $0.028\pm 0.004$ & $0.171\pm 0.012$ & $1.82\pm 0.36$ & $7.09\pm 3.51$ \\
 $0.9<z<1.3$ & 1.02 & 306 & $0.018\pm 0.001$ & $0.140\pm 0.006$ & $1.06\pm 0.44$ & $4.56\pm 1.23$ \\
 $1.3<z<2.3$ & 1.68 & 274 & $0.014\pm 0.001$ & $0.128\pm 0.005$ & $0.12\pm 0.35$ & $3.50\pm 1.33$ \\
\hline
\end{tabular}
\end{table*}

Table~\ref{table:detection:gal} shows
that only a fraction of the initial sample is detected in the far-IR.
To overcome this low detection rate, we can use a stacking method
\citep[e.g.][]{dole2006} to improve our detection threshold.
We start from a catalog of 24~$\mu$m sources
divided into bins of redshift and flux and then stack the corresponding 
70 and 160~$\mu$m images for the sources without direct detections
at these wavelengths. While this yields only the average flux of
the sources, it is very powerful approach if the underlying source
selection at the shorter wavelength is well controlled (e.g. when sources 
belong to small ranges of flux and redshift). 

For the Bo\"otes field we selected galaxies detected at 24~$\mu$m but not
detected at 160~$\mu$m. We used redshift bins of $0<z<0.25$, $0.25<z<0.5$, $0.5<z<1$
and 24~$\mu$m flux bins of and $0.8<S_{24} < 1.5$, $1.5<S_{24}<3$, $3<S_{24}<10$~mJy.
We also used the 291~arcmin$^2$ GOODS/CDFS and GOODS/HDFN fields to extend
the analysis to higher redshifts based on the 24~$\mu$m sample described
by \citet{caputi2007}. This sample consists of 24~$\mu$m selected star-forming
galaxies where AGN have been excluded using both X-ray and near-infrared
(power-law) criteria \citep[see discussion in ][]{caputi2007}. For the
GOODS sample we used redshift bins of 
$0<z<0.3$, $0.3<z<0.6$, $0.6<z<0.9$, $0.9<z<1.3$ and $1.3<z<2.3$
and produced stacked images at 8~$\mu$m, 70~$\mu$m and 160~$\mu$m 
for all sources with $S_{24} > 80 \,\mu\textrm{Jy}$.
We used the photometry methods and calibrations from Sect.~\ref{photometry} 
to measure the fluxes of stacked sources. By design,
we can measure the fluxes in all redshift, flux and wavelength
bins as summarized in Tab.~\ref{table:stack_bootes} for the
Bo\"otes field and Tab.~\ref{table:stack_karina} for the
GOODS fields.

As well as estimating the flux errors as in \S~\ref{photometry}, we also
estimated the uncertainties using a jackknife analysis. Given a sample of 
$N$ sources, we measure the standard deviation of the fluxes found by
stacking many combinations of $N-1$ sources. The standard deviation of
this distribution divided by the square root of the number of stacked sources 
gives the jackknife error bar. In general, the jackknife uncertainties
will be larger than the photometric uncertainties because they also 
include the intrinsic scatter in the fluxes of the stacked population.
Thus, the low ``signal-to-noise'' ratios implied by the jackknife uncertainties 
reported in Tab.~\ref{table:stack_bootes} and Tab.~\ref{table:stack_karina} 
are indicative of significant scatter in the population rather than low
significance in detecting the stacked sources.

Finally, we also compare to the composite SEDs from \citet{zheng2007}, 
who analyzed a sample of 579 optical galaxies ($R < 24$) in the CDFS with 
$0.6 < z < 0.8$ and a stellar mass $M_\star > 10^{10} M_\odot$ based
on a combination of spectroscopic (VVDS \citep{le_fevre2005} and
GOODS \citep{vanzella2005,vanzella2006} surveys) and photometric
(COMBO-17 \citep{wolf2004}) redshifts.
\citet{zheng2007} divided the galaxies with 24~$\mu$m detections
into two bins with equal total 24~$\mu$m  luminosities (the 58
brightest sources in one bin, and the remaining 160 detections
in the second), and then put the remaining 361 galaxies without 
24~$\mu$m detections into a third bin. They then measured the
mean SEDs for the three samples, measuring the 24, 70 and 160~$\mu$m
fluxes by stacking where there were no direct detections.

We will use these stacked ``composite'' sources to confirm that
our low redshift results apply to sources with higher redshifts
and lower infrared luminosities.

\subsection{Summary}

\begin{figure}[!th]
\resizebox{\hsize}{!}{\includegraphics{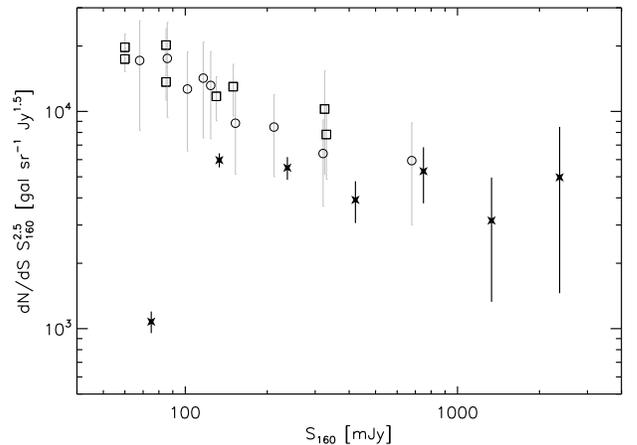}}
\caption{160~$\mu$m differential number counts for our sample of 372 galaxies 
individually detected at 8, 24, 70 and 160~$\mu$m (black stars) as 
compared to the counts obtained in the Marano and CDFS fields by \citet{dole2004}
(open squares) and in the FLS by \citet{frayer2006} (open circles). 
This shows that we are uniformly sampling the far-infrared population over
a large range of 160~$\mu$m fluxes.}
\label{figure:comp160}
\end{figure}

To summarize, our full set of samples consists of:

$\bullet$ First, we have 372 galaxies individually detected 
at 8~$\mu$m, 24~$\mu$m, 70~$\mu$m and 160~$\mu$m. These sources
all have accurate redshifts, based on spectroscopy for AGES and
FLS (93\% of the sources) and COMBO-17 photometric redshifts for the
CDFS (7\% of the sources). While there are some variations in 
the selection criteria for each field, we can view these 
as {\it far-infrared} 160~$\mu$m flux-limited samples.
Figure~\ref{figure:comp160} compares the differential number
counts at 160~$\mu$m for our sample to those from
\citet{dole2004} and \citet{frayer2006} to show that the
completeness of this subsample is $\sim$50\% and that the 
sampling is fairly uniform over a broad range of 160~$\mu$m fluxes.

$\bullet$ Second, we have 13 stacked points constructed from a
sample of $\sim$1700 star-forming galaxies and extending to
redshift $z\sim2$ and 24 $\mu$m fluxes of $S_{24} = 80 \,\mu\textrm{Jy}$ 
that will allow us to probe higher redshifts and lower infrared 
luminosities. These galaxies are typical of {\it mid-infrared}
selected galaxies because they were drawn from a complete sample
selected at 24~$\mu$m.

$\bullet$ Third, we have 3 stacked points from \citet{zheng2007} at 
redshift $\sim 0.7$ that were built from a sample of optically-selected
galaxies.

Thus our combined sample covers a wide range of 24~$\mu$m fluxes and 
redshifts, and is homogeneous over a wide range of 160~$\mu$m fluxes. As it is 
representative of both mid- and far-infrared selected sources, it is well-suited 
for a general statistical study of infrared galaxies, with no obvious biases
towards either cool or warm infrared galaxies.

QSOs have been removed from our sample based on an optical spectroscopic
diagnostic (emission lines). Obviously, not all AGNs have such signatures,
so it is likely that our sample contains some un-identified AGN
\citep[e.g.][]{le_floch2007}. Several criteria based on IRAC colors are 
proposed in the literature to select AGNs \citep[e.g.][]{lacy2004,stern2005,richards2006},
and we note that only $\sim3\%$ (10/372) of our sources lie in the \citet{stern2005}
AGN-selection region. Moreover, \citet{caputi2007} and \citet{fiore2007} have
shown that AGN are a minority of sources (less than 10\% or 5\%, respectively)
at $z \lesssim 1$. Thus, the presence of a few unidentified AGNs will
not affect our lower redshift results ($z\lesssim1.3$). However, contamination
from AGN may not be negligible at higher redshifts 
\citep[e.g.][]{daddi2007a,fiore2007,papovich2007} and the mid-infrared
spectra of $z\sim2$ galaxies may be significantly contaminated by an embedded AGN.
We believe our $z\sim2$ stacking results are little affected by AGNs,  since the
sample was restricted to star-forming galaxies based on many well-defined
criteria \citep{caputi2007}.


\section{Getting the infrared and monochromatic luminosities}

\subsection{A model-independent estimate of the infrared luminosity}
\label{lir}

We define the infrared luminosity as 
\begin{equation}
L_\textrm{IR} = L_{5-1000\mu\textrm{m}} = \int^{1000\mu \textrm{m}}_{5\mu \textrm{m}}
L_\nu \textrm{d}\nu,
\end{equation}
where $L_\nu$ and $L_\textrm{IR}$ are given in W/Hz and W, respectively.
This differs from the definition of $L_{8-1000\mu\textrm{m}}$ as the 
infrared luminosity between 8 and 1000~$\mu$m introduced by \citet{sanders96} 
in order to include the PAH emission in the 5 to 8~$\mu$m range and
to put the wavelength boundary at a more physical frontier between 
stellar and dust emission. For the \citet{lagache2004} templates
for infrared galaxies, we find that the differences between the two
definitions are $(L_{5-1000\mu\textrm{m}}/ L_{8-1000\mu\textrm{m}})= 1.07 \pm 0.04$. 

\begin{figure}[t!]
\resizebox{\hsize}{!}{\includegraphics{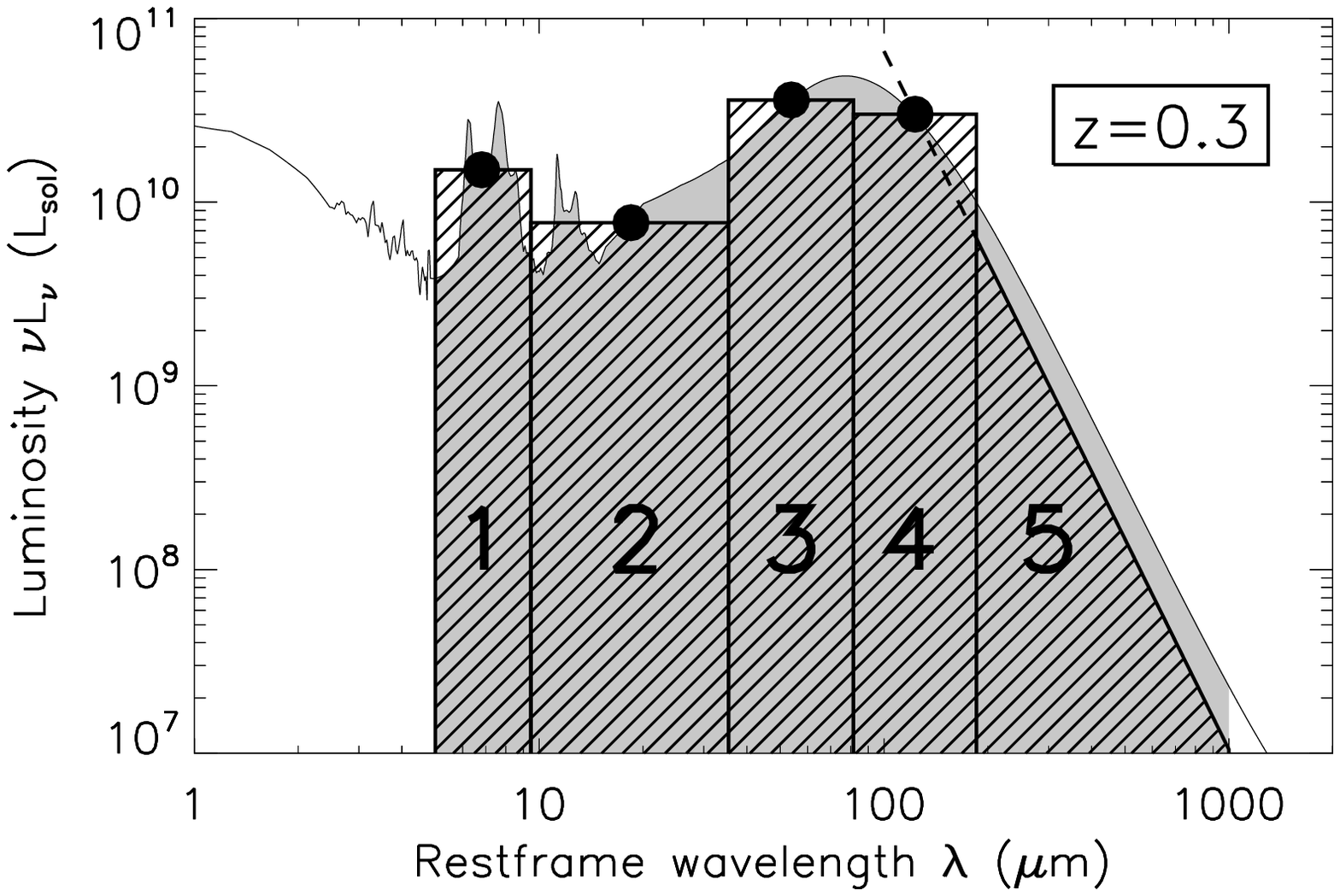}}
\resizebox{\hsize}{!}{\includegraphics{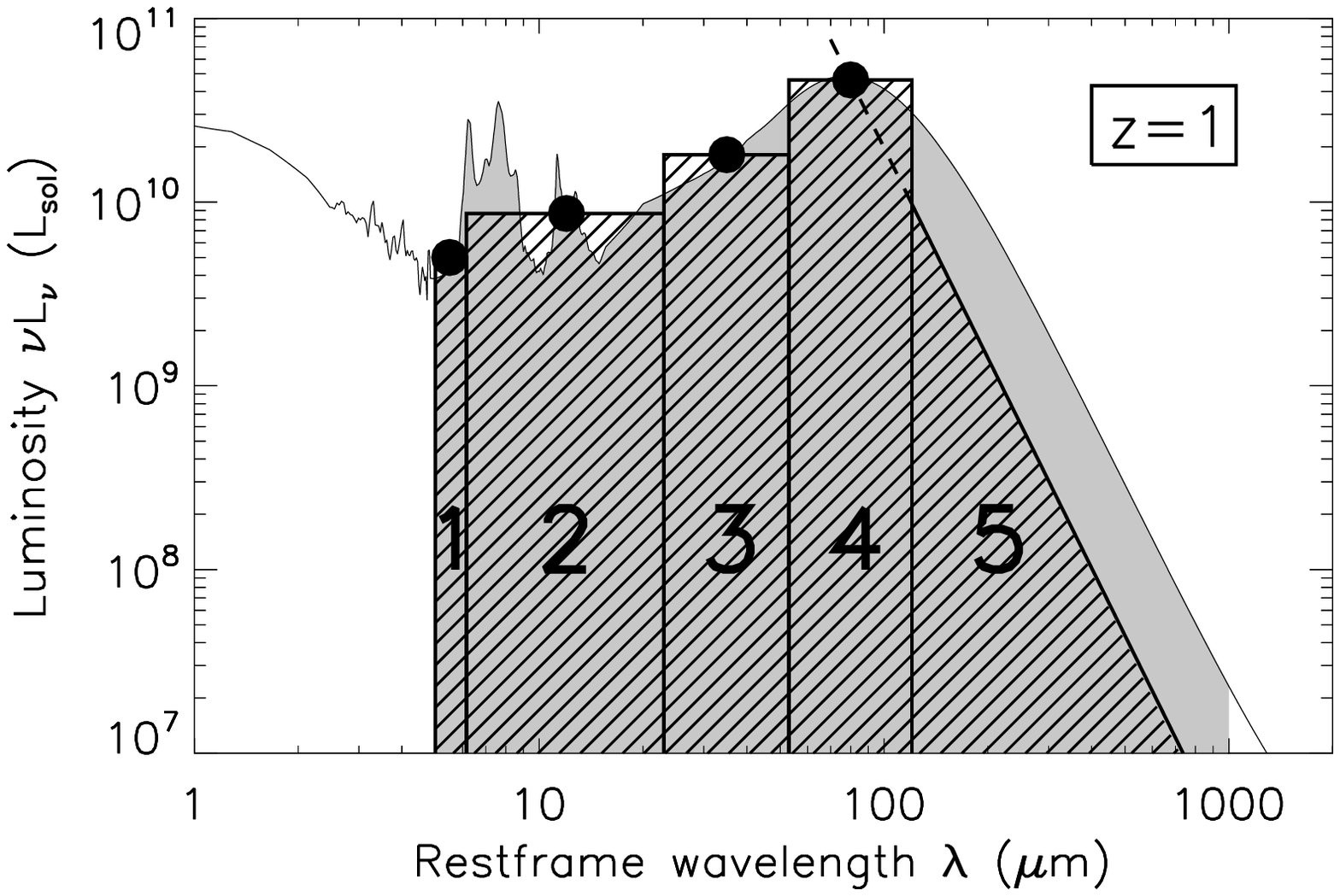}}
\caption{Illustration, for an arbitrarily chosen template, of the method used to 
derive the infrared luminosity from the four observed luminosities
at 8, 24, 70 and 160~$\mu$m and the redshift ({\it top:} $z=0.3$ ;
{\it bottom:} $z=1$). The area of the gray regions is equal to the true
infrared luminosity, while the area of the hatched regions corresponds
to our estimate of $L_\textrm{IR}$. The sizes and positions of the 
five regions are described in the text.}
\label{figure:rectangle}
\end{figure}

We estimated the infrared luminosity from the redshift and the four
observed luminosities at 8, 24, 70 and 160 $\mu$m without fitting
model templates to the data in order to avoid any biases in the 
models such as contamination from AGN or limited ranges of grain
temperatures. Our method simply consists of estimating the total 
infrared flux by adding the luminosities within the 5 regions 
shown in Fig.~\ref{figure:rectangle}. Regions 2, 3 and 4 are 
rectangles centered at 
observed-frame 24, 70 and 160 $\mu$m (i.e. 24/(1+$z$), 70/(1+$z$) and 160/(1+$z$) $\mu$m
rest-frame). Their widths are determined by forcing them to be
contiguous. Region 1 is a rectangle extending from 5~$\mu$m to the beginning
of Region 2. The luminosity at the center of Region 1 is
calculated by linearly interpolating the $\nu L_\nu$ values for 
the observed 8~$\mu$m and 24~$\mu$m points. For $z>1.5$, the width of this 
first region is equal to zero and the observed 8~$\mu$m flux is no 
longer used in the estimate of the infrared luminosity. Lastly, Region 5
is a triangle (on a logarithmic scale) with a slope of $-4$ defined
so that the extrapolation of the edge passes through the observed 
160~$\mu$m point. This slope models the modified black-body emission 
of big grains with a spectral index $\beta=2$ and a dust temperature $T_d = 25 \textrm{ K}$
\citep{sajina2006} or $\beta=1.7$ and $T_d = 35 \textrm{ K}$
\citep{taylor2005} well because the slope between 200~$\mu$m and 1~mm of
these two modified black-body spectra is close to $-4$. 
The slope between 200~$\mu$m and 1~mm measured on different templates
\citep{lagache2004,dale2002,chary2001} varies between $-3.5$ and
$-4$, but varying the slope from $-3.5$ or $-4.5$ has little
($\sim 1\%$) effect on the estimate of $L_\textrm{IR}$.
The rest-frame luminosity $(\nu L_\nu)_\textrm{rest}$ and the observed
luminosity $(\nu L_\nu)_\textrm{obs}$ are then related by
\begin{equation}
(\nu L_\nu)_{\lambda\textrm{,rest}} = (\nu L_\nu)_{\lambda(1+z)\textrm{,obs}},
\end{equation}
where $z$ is the redshift of the source. The statistical uncertainties
in $L_\textrm{IR}$ are easily computed from the uncertainties in the
8, 24, 70 and 160~$\mu$m fluxes since  $L_\textrm{IR}$ is simply a
linear combination of the four bands.

\begin{figure}[t!]
\resizebox{\hsize}{!}{\includegraphics{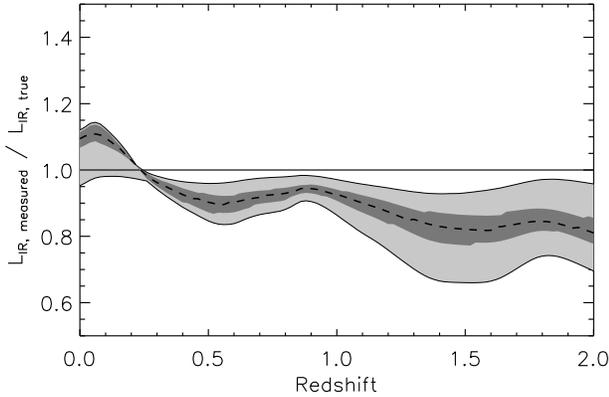}}
\caption{Validation of the infrared luminosity estimate using the
\citet{lagache2004} template
spectra. The shaded bands show the ratio between the estimated and
true values for $L_\textrm{IR}$ as a function of redshift. The light
gray area shows the range encompassing all templates, the dark gray
area shows the range encompassing all templates with ``typical'' luminosities
$-0.5 < \log(\frac{L_\textrm{IR}}{L^\star}) < 0.5$, and the dashed line is
the result for the template with $L_\textrm{IR}=L^\star$. }
\label{figure:compar_rect}
\end{figure}

We tested the method using templates from the \citet{lagache2004} library. We simulated 
observations at 8, 24, 70 and 160 $\mu$m as a function of redshift and compared the 
infrared luminosity obtained with our method to the one obtained by a
proper integration of the template between 5~$\mu$m and 1~mm with the results shown in
Fig.~\ref{figure:compar_rect}. For $0<z<1$, the errors are less than 15\%. 
For redshifts higher than 1, we systematically underestimate the
infrared luminosity and the errors are larger (between 5 and 30\%)
because the peak of the rest-frame FIR emission is leaving
the 160~$\mu$m bandpass as the galaxy redshift increases (Fig.~\ref{figure:rectangle}).
Submillimeter data are needed to better constrain the FIR SED of $z \gtrsim 1.3$ sources.
If we restrict the comparison to templates that are more representative of the typical
luminosities and redshifts of our sources, the errors are smaller. For example, 
if we restrict ourselves to models with $-0.5 < \log(\frac{L_\textrm{IR}}{L^\star}) < 0.5$,
where $L^\star$ is the characteristic luminosity of the infrared luminosity function
from \citet{le_floch2005} (for $0<z<1$) or \citet{caputi2007} (for $1<z<2$), then
the errors are less than 15\% (see the dark area in Fig.~\ref{figure:compar_rect}). 
Although we could try to correct the infrared luminosities for these systematic errors, 
we decided not do so in order to avoid introducing model-dependent correction factors.
In any case, the consequences of this bias are minor and we do not include them
in our error estimates for $L_\textrm{IR}$. We discuss systematic
uncertainties further in Sect.~\ref{Kcorr_eff}.

\subsection{K-correction}
\label{section:kcorr}

We have to apply K-corrections to compute rest-frame luminosities
$(\nu L_\nu)_\textrm{rest}$. We chose to use the \citet{dale2002} model,
which is a physical model based on an incident heating
intensity and a particle size distribution. Unlike other popular, but
more empirical,  models \citep[e.g.][]{chary2001,lagache2004} developed 
for fitting galaxy properties and statistics (like correlations in
monochromatic luminosities, luminosity functions, deep number counts),
the  \citet{dale2002} physical model covers a wide range of dust properties. 
For each galaxy, we search for the template that best fits the four data 
points ($\log(S_8)$, $\log(S_{24})$, $\log(S_{70})$ and $\log(S_{160})$)
and then use it to compute the K-corrections. At higher redshifts,
where observed-frame bands overlap with different rest-frame band,
we use the overlapping bands to compute the K-correction in order 
to minimize the dependence on the templates. For example, at $z=1.7$
we used the observed 24~$\mu$m luminosity to compute the rest-frame
8~$\mu$m luminosity. We use the average redshift and fluxes of the 
stacked sources to compute their K-corrections. In Appendix~\ref{app:kcorr_stacking} 
we show that this approximation is reasonable and that we do not need to
apply K-corrections weighted by the redshift distribution.
The sources that are most affected by the model for the K-corrections 
are those at intermediate redshifts ($z\simeq0.5-0.7$).

\begin{figure}[t!]
\resizebox{\hsize}{!}{\includegraphics{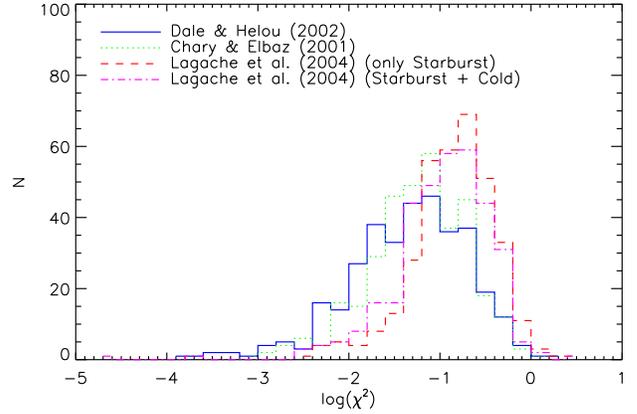}}
\caption{$\chi^2$ histograms obtained with the different template libraries, where
$\chi^2 = \sum_\lambda(\log S_{\lambda, \mathrm{best-fit}} - \log S_\lambda)^2 $.
The solid blue and dotted green lines correspond to the \citet{dale2002} 
and \citet{chary2001} libraries, respectively. The dashed red and dotted-dashed purple 
lines are the distributions for the starburst-only and two-template (starburst $+$ cold)
templates from the \citet{lagache2004} library.}
\label{figure:chi2_library}
\end{figure}

\begin{figure}[!t]
\resizebox{\hsize}{!}{\includegraphics{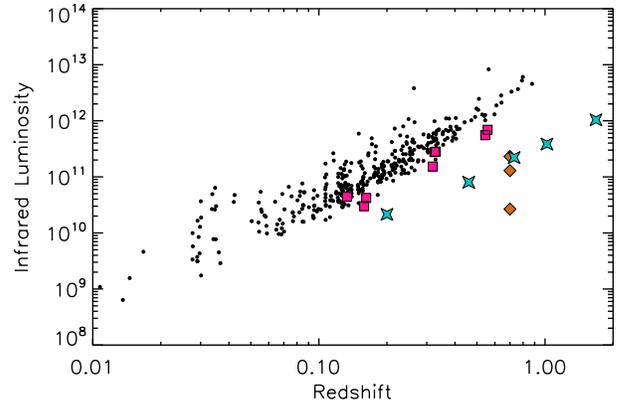}}
\caption{Infrared luminosities as a function of redshift for all
galaxies in our sample. Small black circles correspond to data from the Bo\"otes, FLS and CDFS fields.
The pink squares are the stacking results in Bo\"otes, the 
orange diamonds are the data points from \citet{zheng2007}, and
the blue stars are the stacking results for the \citet{caputi2007}.}
\label{figure:Lbol_z}
\end{figure}

As a consistency check, we also fit the data with the template 
models from \citep{chary2001,lagache2004}. For the \citet{lagache2004} library,
we fit two models, one with only the starburst component and one
which was a linear combination a cold and a
starburst component. In most cases, we find that the
 \citet{dale2002} library produces better fits, as illustrated
in Fig.~\ref{figure:chi2_library}. The medians of the $\chi^2$ distributions 
are 0.052, 0.063, 0.155, 0.128, when using the DH, CE, LDP 
and 2-component LDP models, respectively. Using a Kolmogorov-Smirnov test, 
we checked whether the $\chi^2$ distributions are drawn from the same parent
distribution. At more than 99\% confidence, the CE, the single LDP and the 2-component LDP 
$\chi^2$ distributions are identical, at 94\% confidence the CE distribution is 
compatible with DH, and the DH distribution differs from the single and 2-component LDP
models at 54\% and 76\% confidence, respectively. Since the \citet{lagache2004} templates
were constructed to model galaxy evolution rather than to fit individual spectra,
finding larger $\chi^2$ values when using this library is not very surprising. 
We find small variations in the monochromatic rest-frame luminosities when using
different model families to compute the K-correction.
These variations are smaller than 15\%, 25\%, 20\% and 20\%
at 8~$\mu$m, 24~$\mu$m, 70~$\mu$m and 160~$\mu$m for the full range of templates
and luminosities, and they are smaller than 10\% for galaxies with
$L_{\textrm{IR}} < 10^{11} L_\odot$. While we discuss these questions
further in Sect.~\ref{Kcorr_eff}, such variations have little
effect on the correlations we will be exploring.

\subsection{Characteristics of our sample}

Figure~\ref{figure:Lbol_z} shows the infrared luminosity as a function of redshift
for the 372 individual galaxies in our sample as well as for the 16 points from
the stacking analysis. 
Our sample mostly contains galaxies with infrared luminosities between 
$10^{10} L_\odot$ and $10^{12} L_\odot$, which corresponds to normal star-forming
galaxies and LIRGs. With the stacking results, LIRGs are well sampled up to $z=1.1$.
We directly detect a few ULIRGs up to $z=0.9$, and then the stacking
analysis adds two points at $z=0.9$ and $z=1.7$. Thus, the redshift--infrared
luminosity plane is reasonably well covered by our data.


\section{Correlations between $L_\textrm{IR}$ and $(\nu L_\nu)_\textrm{rest}$ at low and high redshift}

We first present the correlations we obtained for the
individually detected galaxies, and then compare them to 
the correlations obtained after adding the stacking points. 
These correlations then provide useful conversions between
band and total luminosities.

\subsection{Correlations at low redshift}
\label{correl:lowz}

\begin{figure*}[t!]
\resizebox{\hsize}{!}{\includegraphics{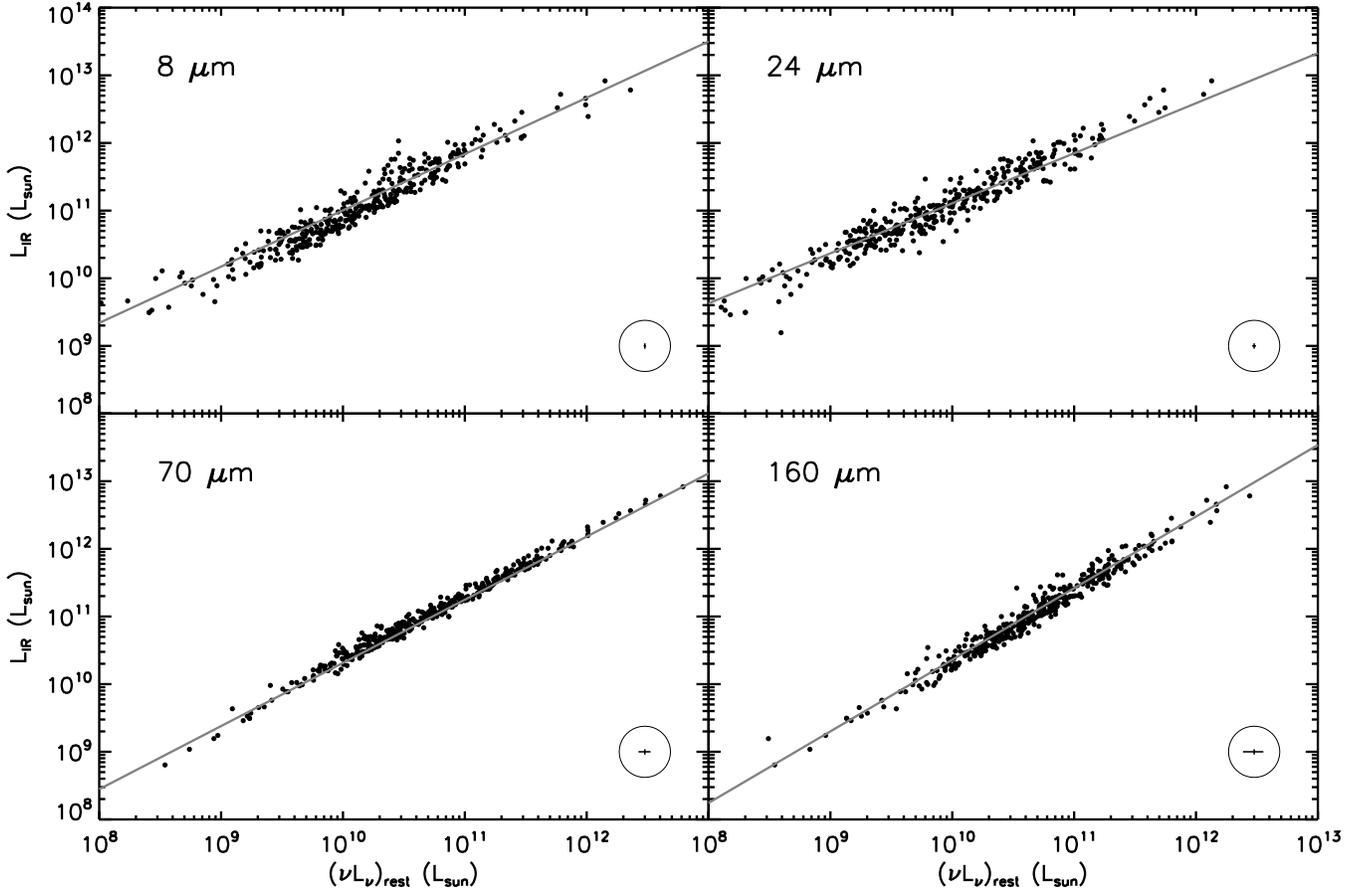}}
\caption{Correlations between rest-frame monochromatic
luminosities and total infrared luminosities at 8, 24, 70 and 160~$\mu$m.
Black filled circles are the data from Bo\"otes, FLS and CDFS fields. The gray lines are
the best fit lines and correspond to Eq.~\ref{eq:corr}. Typical error bars
are shown in a circle in the bottom-right corner.}
\label{figure:correl}
\end{figure*}

We focus on the correlations between the rest-frame monochromatic luminosities 
and the total infrared luminosities, as shown in Fig.~\ref{figure:correl}. To
first order, the correlations are simply a scaling effect -- stronger infrared
emission implies stronger emission at all wavelengths. When we fit the
relationships between  $\nu L_\nu$ and $L_\textrm{IR}$ at the different
wavelengths using linear least-squares fits to the logarithms of the
luminosities and including the uncertainties in both quantities, we 
find that
\begin{equation}
\label{eq:corr}
\left\{
\begin{array}{lclcl}
L_\textrm{IR} & = & 482.5 \times (\nu L_\nu)_{8\mu \textrm{m, rest}}^{0.83}   & & (\pm 37\%) \\
L_\textrm{IR} & = & 5113  \times (\nu L_\nu)_{24\mu \textrm{m, rest}}^{0.74}  & & (\pm 37\%) \\
L_\textrm{IR} & = & 9.48  \times (\nu L_\nu)_{70\mu \textrm{m, rest}}^{0.93}  & & (\pm 16\%) \\
L_\textrm{IR} & = & 0.596 \times (\nu L_\nu)_{160\mu \textrm{m, rest}}^{1.06} & & (\pm 26\%). \\
\end{array}
\right.
\end{equation}
We now see that the logarithmic slopes can differ from unity, which
means that the shapes of the galaxy SEDs depend on luminosity. For example, 
the 24 $\mu$m rest-frame luminosity makes a smaller contribution to the
total infrared luminosity in faint galaxies than in brighter ones.
These relations are illustrated in Fig.~\ref{figure:correl}. 

We also computed the 1-$\sigma$ scatter of the measurements around
the best fitting relations (Eq.~\ref{eq:corr}). These are defined to be the relative uncertainties
in $L_\textrm{IR}$ estimated from
\begin{equation}
\frac{\sigma_{L_\textrm{IR}}}{L_\textrm{IR}} = \ln{10} \times \sigma_{\log{L_\textrm{IR}}}.
\end{equation}
We see that the rest-frame 70~$\mu$m luminosity is the best tracer of the
total infrared luminosity. This means that, for a given 
infrared luminosity, the scatter of $(\nu L_\nu)_{70\mu \textrm{m, rest}}$
is the smallest. This can easily be understood by considering
the two extreme templates shown in Fig.~\ref{figure:simu_templates}.
The two templates are normalized to have the same total infrared 
luminosity, and we see that the rest-frame 70~$\mu$m luminosity
minimizes the scatter because it is close to the intersection of the 
two templates (between 80 and 90 $\mu$m). 
The same argument also explains why the 8 and the 24~$\mu$m luminosities
are the worst tracers of $L_\textrm{IR}$

These correlations were derived based on a far-infrared (160~$\mu$m) selected
sample of galaxies, which will introduce some biases. For general
use, we recommend the more general relations developed in the next
section, even if the differences are small.

\subsection{Validation with higher redshift sources}
\label{correl_highz}

\begin{figure*}[t!]
\resizebox{\hsize}{!}{\includegraphics{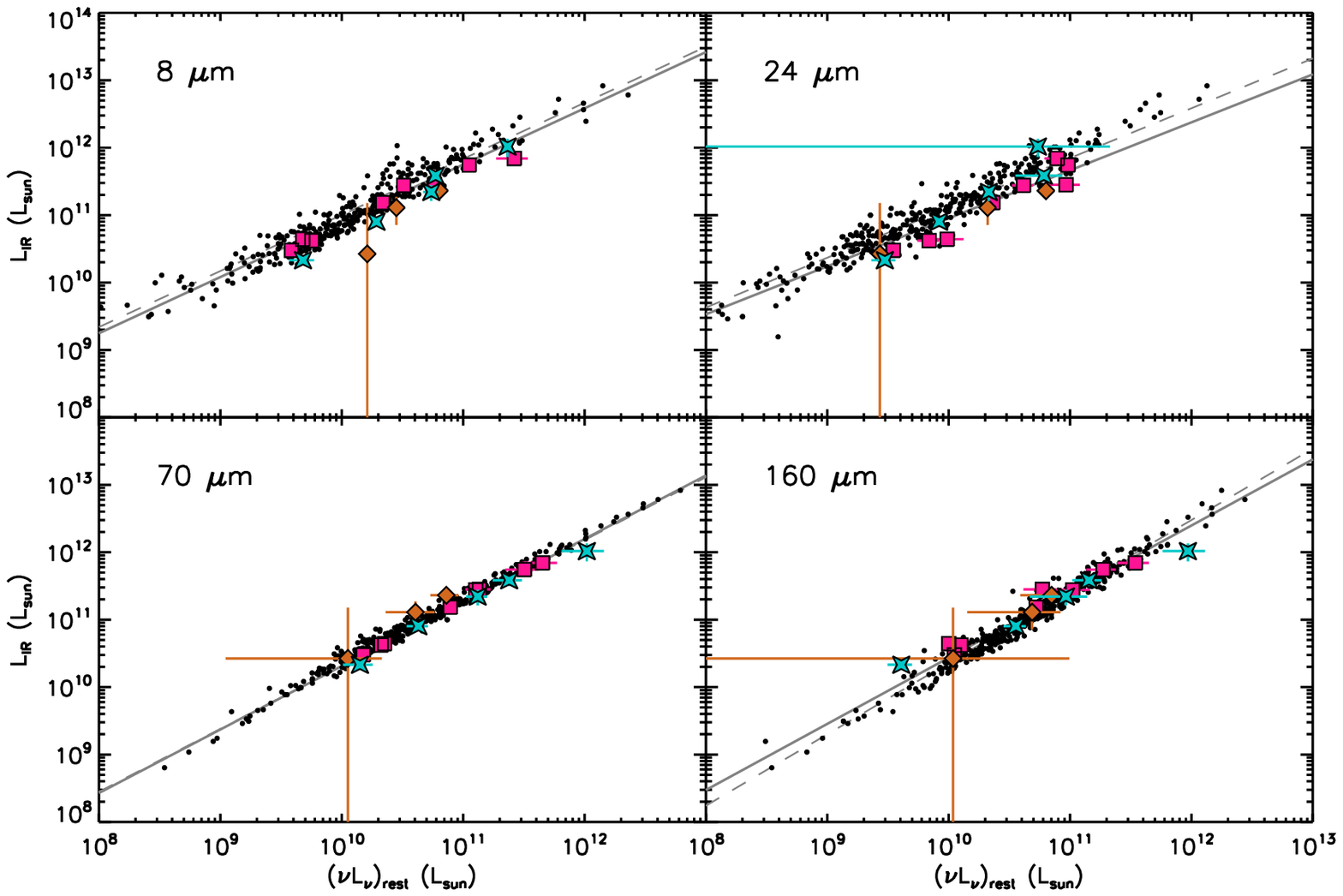}}
\caption{The correlations as in Fig.~\ref{figure:correl}, but with the addition of stacked galaxies.
The symbols have the same definitions as in Fig.~\ref{figure:Lbol_z}.
The gray continuous lines indicate the best fits obtained with all the
data points (Eq.~\ref{eq:corr_new}) and the gray dashed lines correspond to the
best fits obtained without the stacking points (i.e., the same as in
Fig.~\ref{figure:correl} and Eq.~\ref{eq:corr}).}
\label{figure:correl_stacking}
\end{figure*}

The sample from which we derived the correlations Eq.~(\ref{eq:corr})
is dominated by moderate redshift galaxies (93\% lie at $z<0.4$) with
direct far-infrared detections. In order to probe higher redshifts 
for a given infrared luminosity, we make use of the measurements from
our stacking analysis (Fig.~\ref{figure:Lbol_z}). For example,
galaxies with $L_\textrm{IR} = 3 \times 10^{11} L_\odot$ are directly
detected up to $z=0.2$, but the stacking points probe $0.3<z<0.7$.
Fig.~\ref{figure:correl_stacking} shows the luminosity correlations
including both the individually detected galaxies and the stacking
analysis data. While the general agreement is good, there are small
systematic shifts between the two samples. This is probably
explained by selection effects. For example, at a given 24~$\mu$m rest-frame 
luminosity, the stacking method allows us to detect galaxies with lower 
160~$\mu$m luminosities than the direct detections, and thus includes
sources with lower infrared luminosities. Equivalently, galaxies 
selected at 24~$\mu$m are warmer than galaxies selected at 160~$\mu$m.
This hypothesis is confirmed by the simulations presented in 
Appendix~\ref{annex}. 

If we recompute the correlations including the stacking data, giving
each stacking point an additional weight equal to the square root of
the number of stacked sources, we find that 
\begin{equation}
\boxed{
\label{eq:corr_new}
\left\{
\begin{array}{lclcl}
L_\textrm{IR} & = & 377.9 \times (\nu L_\nu)_{8\mu \textrm{m, rest}}^{0.83}    & & (\pm 37\%) \\
L_\textrm{IR} & = & 6856  \times (\nu L_\nu)_{24\mu \textrm{m, rest}}^{0.71}   & & (\pm 54\%) \\
L_\textrm{IR} & = & 7.90  \times (\nu L_\nu)_{70\mu \textrm{m, rest}}^{0.94}   & & (\pm 19\%) \\
L_\textrm{IR} & = & 4.24  \times (\nu L_\nu)_{160\mu \textrm{m, rest}}^{0.98}  & & (\pm 31\%). \\
\end{array}
\right.
}
\end{equation}
These correlations, plotted using solid lines in Fig.~\ref{figure:correl_stacking}, 
show small changes from the results for 160~$\mu$m-selected galaxies in Eq.~(\ref{eq:corr}).
These new relations are 
representative of the total infrared galaxy population, as the full sample includes
all types of galaxies (i.e. warm and cold, corresponding to mid- and far-infrared
selection, respectively) over a broad range of redshifts. In particular, they are reliable up
to $z=1.1$ for LIRGs and $z \sim 2$ for ULIRGs.
 
Note that the scatter in the relations at 24~$\mu$m increased by far more than
the scatter at the other wavelengths. We are uncertain as to the cause.
While we think AGN contamination in the sample is small, it could cause
part of the increase. It may also be due to extra systematic scatter in
the K-corrections for sources at $z  \gtrsim 0.8$ when the PAH features
start to enter the 24~$\mu$m band.

\subsection{Uncertainties from the integration
of the SED and the K-correction}
\label{Kcorr_eff}

The systematic uncertainties in these relations arise from uncertainties
in our estimate of $L_\textrm{IR}$ and any errors in the K-corrections.

We showed in Fig.~\ref{figure:compar_rect} that the errors in the
estimate of $L_\textrm{IR}$ can be quite large at high redshift (for
example 30\% at redshift $z=1.5$). But these large errors are only
found for the coldest (i.e. less luminous) templates of the \citet{lagache2004} 
library and these quiescent galaxies do not seem to be representative of the
distant universe. More realistic templates, with a typical luminosity
of $L^\star$ from observed luminosity functions \citep{le_floch2005,caputi2007}
at these redshifts show much smaller errors (about 15\%, see the dark area
on Fig.~\ref{figure:compar_rect}). Thus, we estimate that the systematic
uncertainties in $L_\textrm{IR}$ are less than 15\% for redshifts $<1$ and 
less than 20\% for redshifts up to $z=2$. As most of the galaxies in our 
sample lie at $z<1$, such small uncertainties will have little effect on 
the estimated correlations.

\begin{table}[t!]
\caption{Changes in the correlations obtained for typical starburst, LIRGs and 
ULIRGs when using the \citet{chary2001} model instead of the \citet{dale2002}
templates for the K-corrections.}
\label{table:error_Kcorr}
\centering
\begin{tabular}{c c c c c}
\hline\hline
$L_\textrm{IR}$  &  8~$\mu$m & 24~$\mu$m & 70~$\mu$m & 160~$\mu$m \\
\hline
$3 \times 10^{10}$ & $+ 3\%$ & $-6\%$ & $+5\%$ & $-11\%$ \\
$3 \times 10^{11}$ & $+ 8\%$ & $-6\%$ & $+3\%$ & $- 4\%$ \\
$3 \times 10^{12}$ & $+13\%$ & $-6\%$ & $<1\%$ & $+ 3\%$ \\
\hline
\end{tabular}
\end{table}

To obtain the rest-frame luminosities for each band, we have to compute and apply 
a K-correction, which is derived from template fits to the data. The choice
of the model was already seen in \S3.2 to have little effect (less than 25\% for 
$L_\textrm{IR} > 10^{11} L_\odot$ and less than 10\% for 
$L_\textrm{IR} < 10^{11} L_\odot$). Tab.~\ref{table:error_Kcorr} shows
the luminosity-dependent changes in the correlations if we use the
\citet{chary2001} models instead of the \citet{dale2002} models to
compute the K-correction. Systematic uncertainties found for typical
starbursts, LIRGs and ULIRGs are generally less than 10\% at 24,
70 and 160~$\mu$m and modestly larger (15\%) at 8~$\mu$m.

\subsection{Useful relations to estimate $L_\textrm{IR}$}

In the last few sections we have shown that the knowledge of one
infrared flux between 8 and 160~$\mu$m can provide a very reasonable
estimate of the total infrared luminosity and hence of the star 
formation rate. In this section we explore whether combining
several monochromatic luminosities can significantly improve the
estimates. We fit the infrared luminosity as a sum of power-law
relations for each wavelength,
\begin{equation}
\label{eq:formula_fit}
L_\textrm{IR} = \sum_{\lambda=8, 24, 70, 160 \, \mu \textrm{m}} 
a_\lambda (\nu L_\nu)_{\lambda, \textrm{rest}}^{\beta_\lambda},
\end{equation}
where the coefficients $a_\lambda$ and slopes $\beta_\lambda$ are free 
parameters. We fit the data, including the stacking results, using
combinations of two wavelengths, three wavelengths or all four 
wavelengths, as summarized in Tab.~\ref{table:formules}.

\begin{table*}[t!]
\caption{Results of the different fits (see text and
Eq.~\ref{eq:formula_fit}) on the whole sample (the directly detected
galaxies plus the stacking points). An empty case means
that the given $a_\lambda$ was fixed to zero. For comparison, we report the
previous relations given in Eq.~\ref{eq:corr_new}.}
\label{table:formules}
\centering
\begin{tabular}{c | c c | c c | c c | c c | c}
\hline\hline
Number of bands & $a_8$   &  $\beta_8$ &  $a_{24}$  &  $\beta_{24}$ &  
$a_{70}$  & $\beta_{70}$ &  $a_{160}$  & $\beta_{160}$  &  1-$\sigma$ \\
\hline
1 band & 377.9  & 0.83 &         &      &        &      &         &      &   37 \% \\
       &        &      &  6856   & 0.71 &        &      &         &      &   54 \% \\
       &        &      &         &      &   7.90 & 0.94 &         &      &   19 \% \\
       &        &      &         &      &        &      & 4.24    & 0.98 &   31 \% \\
\hline
2 bands & 5607   & 0.71 & $1.00 \times 10^{-5}$   & 1.50 &       &      &       &      &   34 \% \\
        &        &      & $8.9 \times 10^{-4}$ & 1.27 &  12.62 & 0.92 &       &      &   20 \% \\
        &        &      &         &      &  3.87  & 0.96 &  1.58 & 0.95 &   11 \% \\
\hline
3 bands & 0.0071  & 1.11 & $7.4 \times 10^{-4}$   & 1.28 & 12.8 & 0.92 &       &      &   19 \% \\
        &        &      &  1.62   & 0.99 &  1.59  & 0.98 &  3.78 & 0.94 &    7 \% \\
\hline
4 bands &  8.86  & 0.81 &  1.28   & 1.00 &  1.45  & 0.98 &  3.92 & 0.94 &    6 \% \\
\hline
\end{tabular}
\end{table*}

Obviously, when we use more than one monochromatic luminosity, we get 
a more accurate estimate of $L_\textrm{IR}$. For example, estimating
the infrared luminosity from the 8 and 24~$\mu$m luminosities leads
to a scatter about the resulting correlation of only 34\% instead
of the 37\% and 54\% found for the individual luminosities. Using
the three far-infrared bands or all four bands give very accurate
results, scatters of 7\% and 6\% respectively, both because the 
infrared emission is dominated by emission from large grains that
peaks between 80 and 150~$\mu$m and becuase these linear combinations
can closely approximate our method for constructing $L_\textrm{IR}$
from the data. Once the scatter is significantly smaller than 
$\sim 25\%$, the uncertainties are dominated by systematic errors.

Such empirical relations may be very useful in practice for measuring 
the total infrared luminosity of infrared star-forming galaxies from 
limited data. In particular, mid-infrared fluxes (8 and 24~$\mu$m) are 
particularly easy to obtain for large numbers of sources and are
well-suited for estimating $L_\textrm{IR}$ and conducting statistical
studies of star formation in LIRGs and ULIRGs. Moreover, the estimates
are nearly independent of the choice of a model, so it is easy to 
obtain relatively precise estimates for the infrared luminosity of 
starburst galaxies (30\% 1-$\sigma$) without any strong assumptions.

\subsection{Comparison with previous studies}
\label{par:compar_other}

\begin{figure}[t!]
\resizebox{\hsize}{!}{\includegraphics{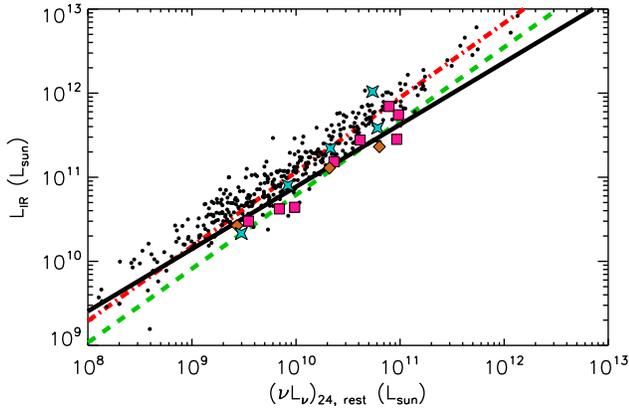}}
\caption{Correlation between $(\nu L_\nu)_{24\mu \textrm{m, rest}}$
and $L_\textrm{IR}$ for our whole sample (same symbols as
in Fig.~\ref{figure:correl_stacking}). The best fit
(Eq.~\ref{eq:corr_new}) is shown with a black solid line.
The green short-dashed line is the relation from \citet{takeuchi2005a} 
and the red dotted-dashed line is the relation from \citet{sajina2006}.}
\label{figure:compar_other}
\end{figure}

Figure~\ref{figure:compar_other} compares our correlation (Eq.~\ref{eq:corr_new})
between the  24~$\mu$m luminosity and $L_\textrm{IR}$ to earlier results
from \citet{sajina2006} and \citet{takeuchi2005a}.
The \citet{sajina2006} sample consists of ISO FIRBACK sources selected at
170~$\mu$m and the \citet{takeuchi2005a} sample was mainly selected at
100~$\mu$m because they required sources to be detected in all four IRAS
bands (12, 25, 60 and 100~$\mu$m). As suggested by \citet{sajina2006},
the difference between these two results can be attributed to selection effect, 
with IRAS-selected sources being warmer than ISO sources. This effect
is similar to the one we discussed in Sect.~\ref{correl_highz} and provide
details for in Appendix~\ref{annex}. Our correlation is roughly bounded by
the relations from \citet{takeuchi2005a} and \citet{sajina2006} -- the warmest galaxies 
in our sample (the stacking points) agree well with the relation of
\citet{takeuchi2005a}, while the coldest points are in better agreement
with the prediction of \citet{sajina2006}. We are sampling a wider range of 
temperature than these previous studies because of our broader selection
criteria. Finally, \citet{dale2002} also tried to estimate the total infrared
luminosity based on a linear combination of the three MIPS monochromatic
luminosities. If we estimate $L_\textrm{IR}$ using their relations we
find good agreement, with a systematic shift of only 6\% and an rms scatter
of 6\%.

\section{Evolution of galaxies and application to the High-Redshift Universe}

\begin{figure*}[t!]
\resizebox{\hsize}{!}{\includegraphics{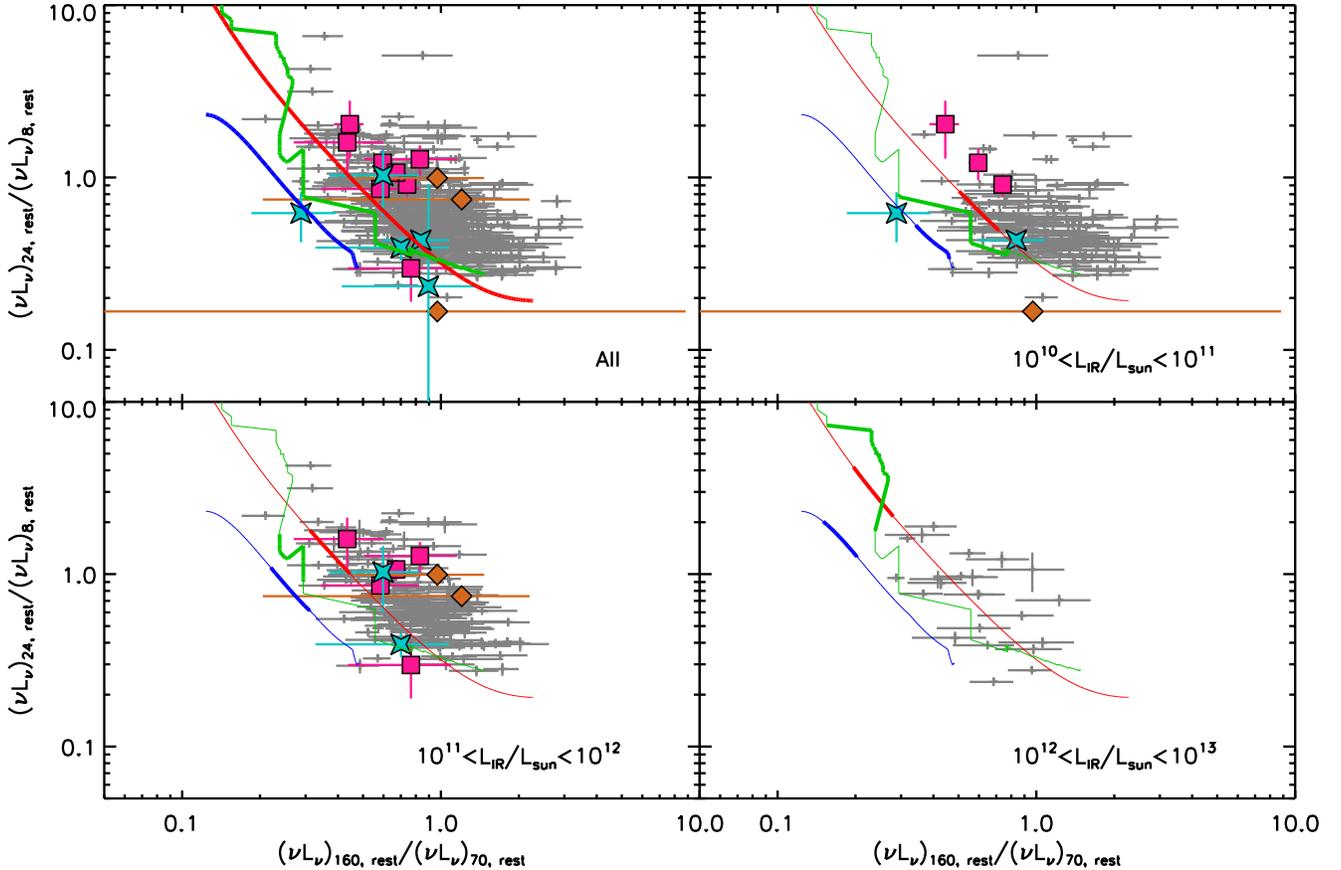}}
\caption{Infrared colors of the galaxies in our sample using the same symbols as
in Fig.~\ref{figure:correl_stacking}. The upper left panel shows all 
luminosities and the other panels show different ranges of infrared luminosity 
$L_\textrm{IR}$. The predictions of the \citet{lagache2004}, \citet{chary2001} and
 \citet{dale2002} models are shown by the blue, green 
and red lines, respectively. In each case, the thick part of each line
corresponds to the luminosity range considered in the panel.}
\label{figure:evol_color}
\end{figure*}

\subsection{Application to the High-Redshift Universe}

Perfect measurements of the infrared luminosity of high redshift galaxies 
requires well-sampled, rest-frame infrared SEDs. Unfortunately, obtaining such 
data is very observing-time consuming because of the relatively low sensitivity
of far-infrared data. Moreover, at high redshift, the maximum of the infrared
emission which is due to emission by big grains is redshifted to submillimeter 
wavelengths. In order to study star formation at high redshift, we need to estimate
the total infrared luminosity with as few parameters as possible. Our study
shows that the total infrared luminosity can be well constrained from the
8 or 24~$\mu$m rest-frame luminosities (with uncertainties of 37 and 54\% 
respectively) and that combining these two luminosities gives a modestly
better estimate (uncertainties of 34\%). \citet{caputi2007} used our conversion 
between the rest-frame 8~$\mu$m luminosity and the $L_\textrm{IR}$ to determine 
the bolometric infrared luminosity function at $z\sim2$. They show that 90\% of 
the infrared energy density due to $z \sim 2$ star-forming systems is produced 
equally by LIRGs and ULIRGs, while LIRGS dominate the emission at $z\sim1$.
A more accurate estimate of 
$L_\textrm{IR}$ can be obtained given the 70~$\mu$m luminosity, with a
scatter of only $19\%$ (1-$\sigma$). It will be very interesting to test
whether these relations hold for the individually detected, faint far-IR
sources that will be found in ongoing ultra-deep 70~$\mu$m surveys \citep[e.g.][]{frayer2006}.

\subsection{Evolution of SEDs?}

As our galaxies span a wide range of infrared luminosities and redshifts, it is 
interesting to investigate whether we observe any evolution within our sample. 

\begin{figure}[t!]
\resizebox{\hsize}{!}{\includegraphics{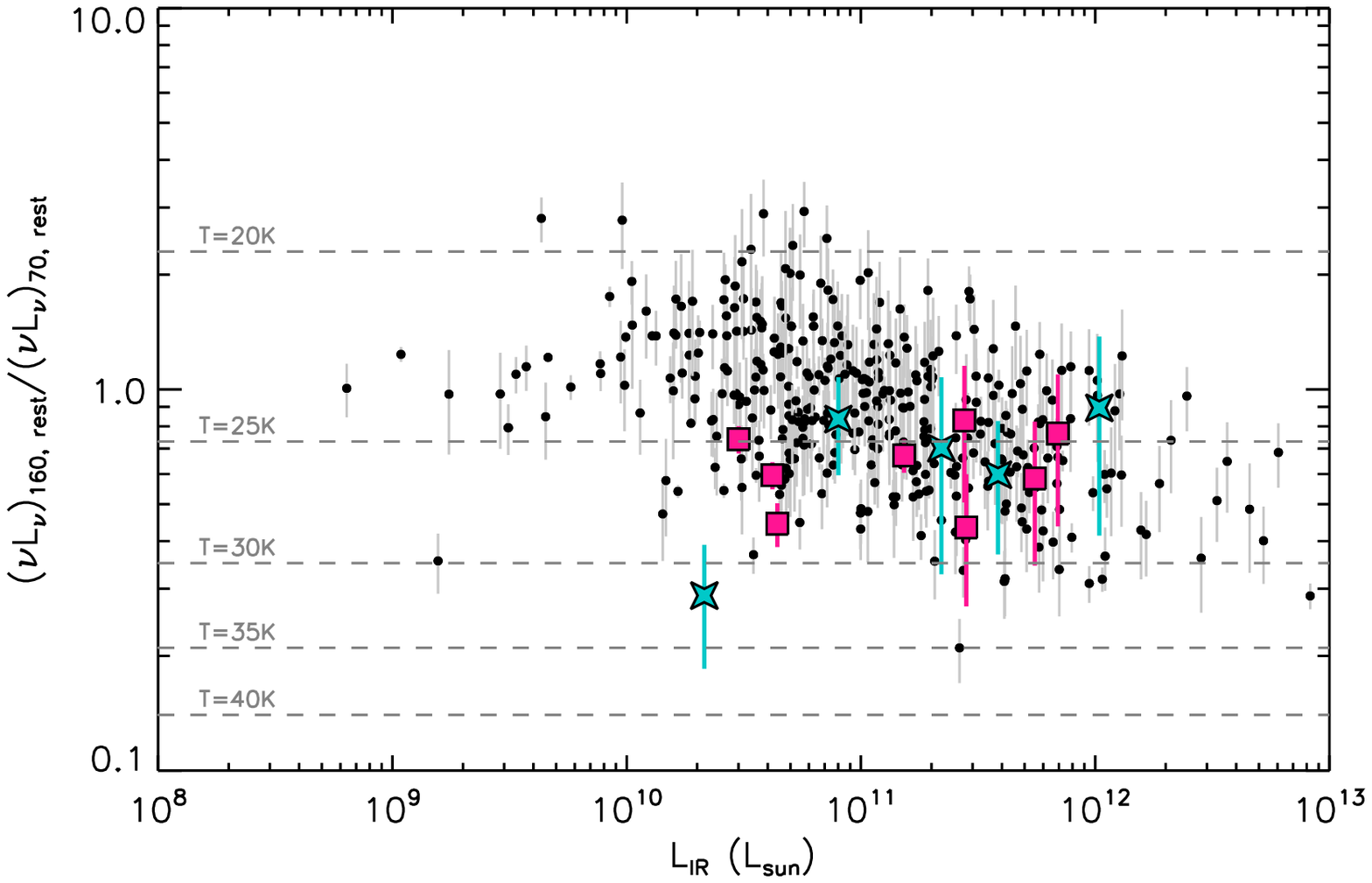}}
\caption{Evolution of the relationship between
$(\nu L_\nu)_{160\mu \textrm{m, rest}}/(\nu L_\nu)_{70\mu \textrm{m, rest}}$ 
and $L_\textrm{IR}$ with redshift where the black circles correspond to
the low redshift sample, the pink squares to intermediate redshifts and 
the blue stars to high redshifts using the same symbols as in Fig.~\ref{figure:Lbol_z}. 
The correspondence between the rest-frame 160/70
color and the big grain temperature is shown by the gray dashed lines assuming
a modified black-body spectrum with a spectral index $\beta = 2$.}
\label{figure:evol_temp}
\end{figure}

Figure~\ref{figure:evol_color} compares the rest-frame 24/8 and 160/70
infrared colors of our low redshift, directly detected galaxies
to those of the stacked galaxies at higher redshifts.
The $(\nu L_\nu)_{24\mu \textrm{m, rest}}/(\nu L_\nu)_{8\mu \textrm{m, rest}}$
color traces the balance between PAHs and Very Small Grains (VSGs), while
the $(\nu L_\nu)_{160\mu \textrm{m, rest}}/(\nu L_\nu)_{70\mu \textrm{m, rest}}$
color is set by the temperature of the big grains. In this rest-frame
color-color diagram, all the high redshift points are compatible
with the lowest redshift sources, which suggests that there is little 
evolution in the dust content of infrared galaxies between $z \sim 0.16$
(which is the median redshift of our sample of directly detected galaxies)
and $z \sim 1.5$. If we compare the data to the predictions
of several SED models \citep{lagache2004,dale2002,chary2001}, we find
the templates of \citet{dale2002} show the best agreement with the data.
The \citet{lagache2004} starburst model underestimates both colors, probably
because it over estimates the dust temperature, and the
\citet{chary2001} model gives intermediate results. These differences
are related to the discussions of these templates in 
Sect.~\ref{section:kcorr} and Fig.~\ref{figure:chi2_library}.
The differences are not a consequence of our template choice -- computing
the K-corrections using the other templates produces similar rest frame
colors for the data and similar levels of agreement between the models
and the data (see Appendix~\ref{app:kcorr_model}).

\subsection{Far-infrared vs Submillimeter galaxies}

\begin{figure}[!ht]
\resizebox{\hsize}{!}{\includegraphics{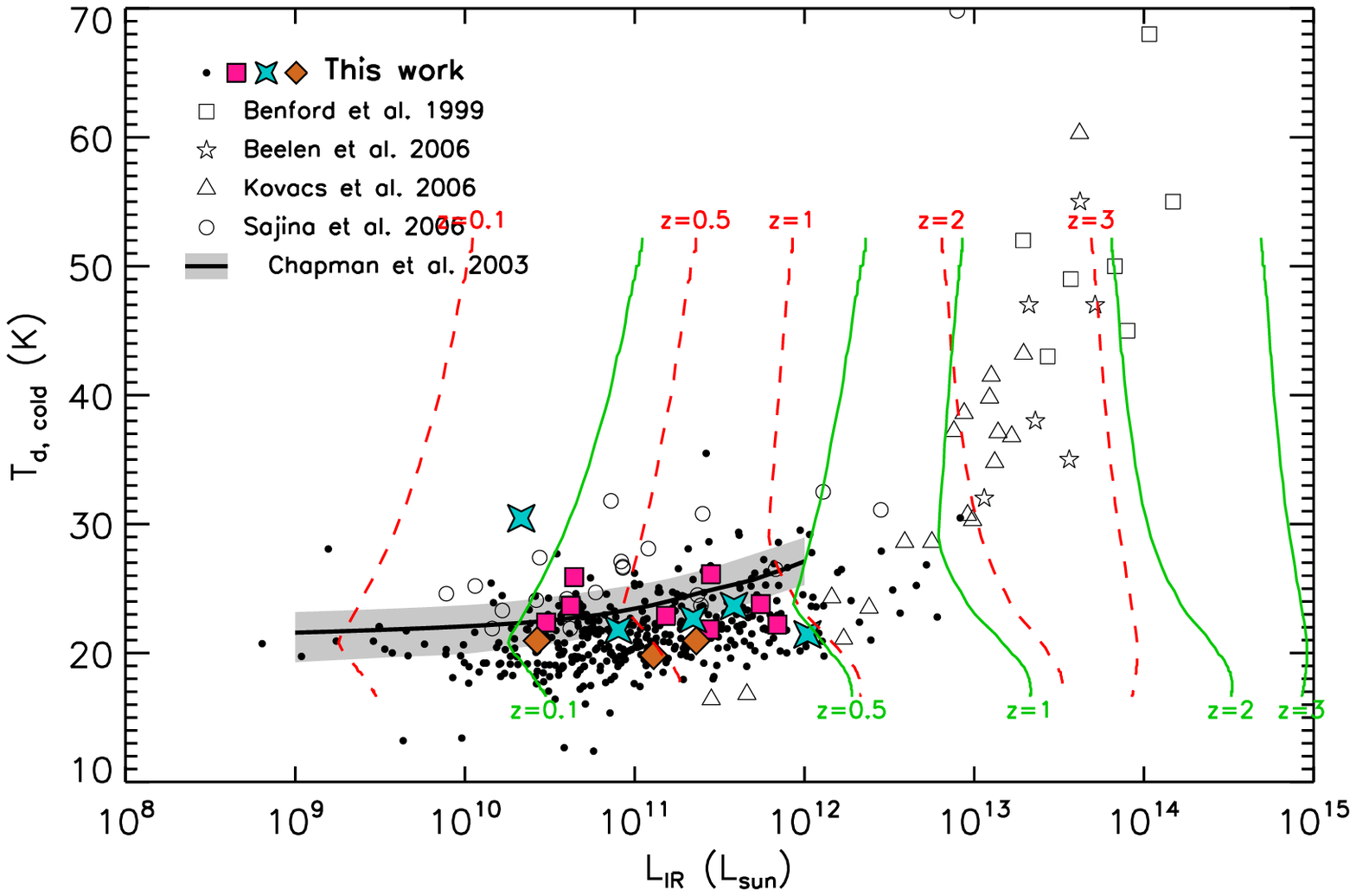}}
\resizebox{\hsize}{!}{\includegraphics{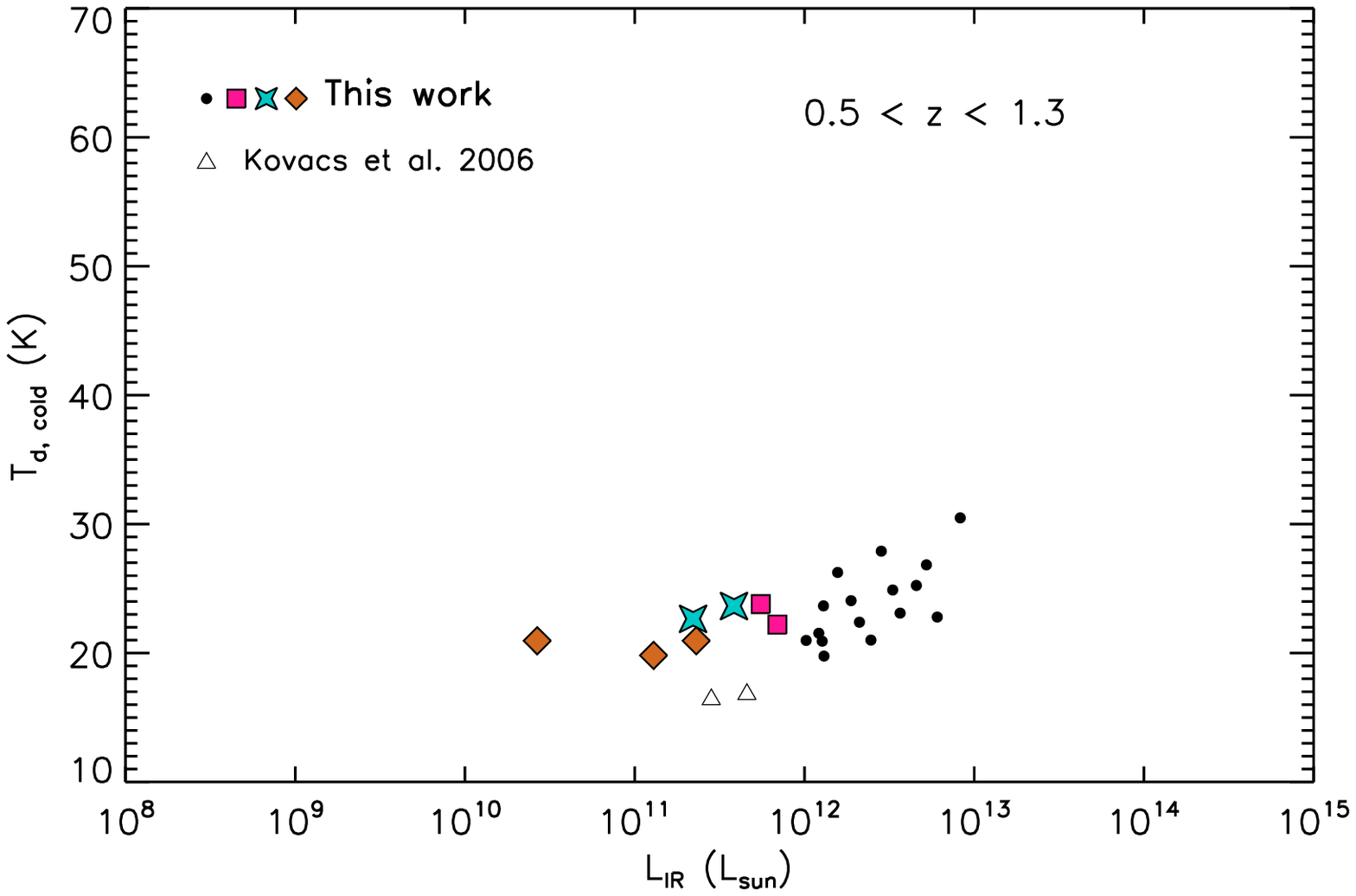}}
\resizebox{\hsize}{!}{\includegraphics{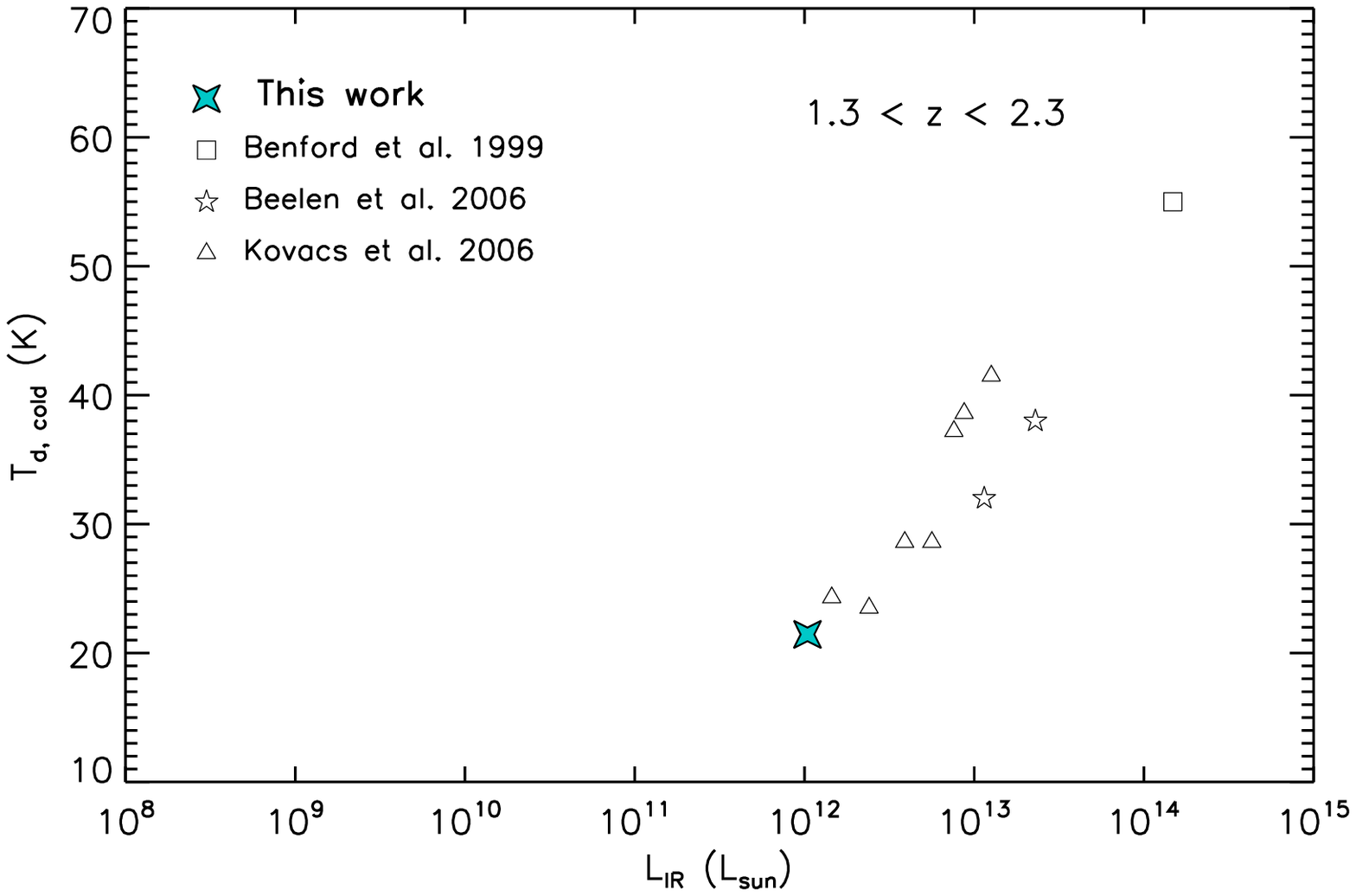}}
\caption{{\it Top:} Comparison of our sample with previous studies using
the same symbols as in Fig.~\ref{figure:Lbol_z}. The thick line is
the relation found by \citet{chapman2003} for local IRAS sources with 
the 1$\sigma$ scatter around their best fit indicated by the light gray
band. The open circles are from the sample of FIRBACK sources \citep{sajina2006},
the open triangles are the $1.5<z<3.5$ SMGs from \citet{kovacs2006}, and the
dusty quasars from \citet{benford99} and \citet{beelen2006} are marked
by the open squares and open stars, respectively. The solid cyan and
dashed orange lines show the detection limits for the individually detected
and stacked sources in the Bo\"otes field as a function of
redshift. Only sources on the {\it right side of the lines}
are detectable. The lower panels are the same, but we restrict the
comparisons to sources with $0.5 < z < 1.3$ ({\it Middle}) and 
$1.3 < z < 2.3$ ({\it Bottom}). }
\label{figure:compar_temp_lir}
\end{figure}

Figure~\ref{figure:evol_color} also shows that the 160/70 color decreases with increasing
galaxy luminosity, which means that the brightest galaxies are the warmest.
This well known property of infrared galaxies \citep[e.g.][]{soifer87} is
also reproduced by the models.

In Fig.~\ref{figure:evol_temp} we examine the evolution of the relationship 
between dust temperature and $L_\textrm{IR}$ with redshift. We have, effectively,
three redshift bins: a low redshift sample made up of the 372 directly detected 
galaxies, an intermediate redshift sample corresponding to the 8 stacking
points from the Bo\"otes field (red squares) and the high  redshift
sample consisting of the 5 stacking points from the CDFS and HDFN
(purple stars). Figure~\ref{figure:evol_temp} shows that there is no
significant evolution in the relationship between $L_\textrm{IR}$ and the dust
temperature with redshift. All the data lie in the same region, even if
we observe some small differences between the stacking points and the
directly detected galaxies. At low luminosity ($L_\textrm{IR} < 10^{11} L_\odot$),
the stacking points are warmer than the sources directly detected in all bands, 
while the reverse is seen at higher luminosities. This should be interpreted as
a selection effect, as we confirm with the simulations presented in 
Appendix~\ref{annex}.

We can also compare the relation between dust temperature and total infrared luminosity
with other published samples of SMGs and AGNs (Fig.~\ref{figure:compar_temp_lir}).
Radio and submillimeter data probe the coldest dust component. In order
to compare our results with longer wavelength surveys, we have to
estimate the temperature $T_{d\textrm{, cold}}$ of the cold dust
component. We used the \citet{dale2002} library to convert the far-infrared
colors (100/60 and 160/70) into estimates of $T_{d\textrm{, cold}}$ by 
fitting a modified blackbody to the templates between
100 and 500~$\mu$m. We find typical temperatures in the range of 18-30~K.
\citet{yang2007} and \citet{yang2007a} found higher dust temperatures
for the same range of infrared luminosities for sources with $0<z<1$. 
Their dust temperatures were determined by fitting modified black-body 
spectra to at least three photometric points between 60 and 850~$\mu$m.
We suspect they find higher temperatures because for $z>0.2$ their
60~$\mu$m fluxes will be contaminated by VSG emission that biases
the temperature upwards. As shown in \citet{yang2007}, the
difference can also be attributed to different spatial scales of
the star formation process, where higher temperatures could indicate that
the star formation occurs in more concentrated regions.
Figure~\ref{figure:compar_temp_lir} shows the relation between
our estimates for $T_{d\textrm{, cold}}$ and $L_\textrm{IR}$. 
We also include the data from \citet{chapman2003} for local IRAS sources,
where we used the same method to convert the 100/60 colors
into cold dust temperatures. Their results are in good agreement 
with our own. We also included data from
\citet{sajina2006} (FIRBACK 170~$\mu$m sources),
\citet{kovacs2006} (SMGs),
and \citet{benford99,beelen2006} (dusty quasars).

Sub-millimeter surveys do not detect warm or low luminosity sources
\citep{chapman2005}, so the infrared and sub-millimeter analyses seem
to be complementary, and by comparing the two approaches we can study
whether the SMGs are a dominant or marginal population for galaxy
evolution. From the previous samples we selected sub-samples 
consisting of the infrared galaxies in SMGs with $0.6<z<1.3$ 
and $1.3<z<2.3$. These sub-samples are shown in separate panels
of Fig.~\ref{figure:compar_temp_lir}. At $z\sim1$, we see that the 
two SMGs are colder than our infrared sources and that we do not detect
a large population of such cold galaxies in the infrared. A small
fraction of cold sources could exist and would be folded into our
stacking points. At $z\sim2$, our stacking point is in good agreement with the 
lowest luminosity SMGs. Unfortunately, the GOODS fields we used to
build our $z\sim 2$ sample are too small to include any of the 
higher luminosity galaxies. However, the good
agreement with SMGs where they do overlap suggests that high-redshift SMGs
are similar to infrared star-forming galaxies. This suggests that
infrared and submillimeter/radio surveys are exploring the same
source population but the methods compliment each other because the 
infrared is well suited for $z\lesssim2$ sources, while the
submillimeter is better for high-redshift sources
because of the advantageous K-corrections in this wavelength range.


\section{Conclusions}

In this paper, we have presented correlations between rest-frame 8~$\mu$m,
24~$\mu$m, 70~$\mu$m and 160~$\mu$m luminosities and an estimate for the
 total infrared luminosity derived without making any assumption on the
shape of the SED that might bias the results. For a sample of 
372 far-infrared (160~$\mu$m)-selected galaxies with $z<0.8$ we
found that the infrared monochromatic luminosities are strongly correlated 
with the total infrared luminosity $L_\textrm{IR}$ and we derived
relations to estimate $L_\textrm{IR}$ from the monochromatic luminosities. In order to
validate this result at higher redshifts, we used a stacking analysis to extend
the data to fainter and higher redshift galaxies. For $z<2$ galaxies selected
at 24~$\mu$m the new data agrees well with the local sample up to a small
systematic shift that we attribute to the differences in the selection criteria 
 -- on average, galaxies selected in the mid-infrared are warmer than those
selected in the far-infrared. The revised correlations including both
samples are probably better for general use. As expected we find that 
combining several monochromatic infrared luminosities yields a more 
precise estimate of the total infrared luminosity than using a single 
luminosity. Since the correlations were derived from a large number of
galaxies with a wide range of luminosities and temperatures and extending
to $z\sim 2$, they should hold for most star-forming galaxies.
In particular, they are applicable for LIRGs up to $z\sim1.1$ and for 
ULIRGs up to $z\sim2$. Extrapolations to higher redshifts, although not 
tested here, should give reasonable results. It is important to remember 
that all known QSOs were removed from our sample. While a similar study
of AGN sample would be very interesting, we have only 7 QSOs detected in
all four bads (8, 24, 70, 160~$\mu$m) and cannot carry out the analysis.
Our correlations should not be used for AGNs unless further tests
demonstrate their validity.

\citet{dale2005} claim that it is dangerous to use the 8~$\mu$m luminosity
as a tracer of the total infrared luminosity because they observed strong
variations (about a factor 10) of this ratio for their sample of
nearby galaxies (SINGS). However, their sample contains many different
objects with very different dust properties and they examined different
regions inside galaxies in detail. Our study shows that at higher redshift 
and on larger scales, the integrated galaxy and dust properties of star-forming
galaxies are more homogeneous, and that using the 8 or 24~$\mu$m luminosities to 
estimate the total infrared luminosity has uncertainties of between 40--50\% that
are much smaller than those given by \citet{dale2005}.
We also compared our results at 24~$\mu$m to previous studies and
found good agreement. The differences we observed can be explained
by differences in the sample selection criteria.

For the sample as a whole, we find no evidence for significant evolution in
the far-infrared SED properties of infrared galaxies with redshift. Both 
the infrared colors and the relationship between dust temperature and $L_\textrm{IR}$ 
of high redshift galaxies from the stacking analysis are compatible with the 
galaxies in the low redshift sample. A small evolution amount
is not detectable because we used different selection criteria at low
(far-infrared) and high (mid-infrared) redshifts. Finally, we
compared our sample to submillimeter data and found that the cold
SMGs observed at $z\sim1$ are a marginal population that is not 
representative of infrared star-forming galaxies. The infrared is the most powerful
wavelength range to study the evolution of star-forming galaxies at
$z\lesssim2$ because submillimeter surveys only select the coldest galaxies.
However, submillimeter and radio wavelengths are more powerful for
higher redshift, $z\gtrsim2$ dusty galaxies, because they better probe 
the dust emission peak. Thus the two approaches complement each other
for studies of galaxy dust properties over cosmic time.

\begin{acknowledgements}
We are grateful to the anonymous referee for a careful reading of the manuscript.
We also thank David Elbaz, Alexandre Beelen and Delphine Marcillac for fruitful discussions.
This work is based on observations made with the {\it Spitzer} Space Telescope,
which is operated by the Jet Propulsion Laboratory, California Institute
of Technology under a contract with NASA. 
\end{acknowledgements}

\bibliographystyle{aa}

\begin{thebibliography}{53}
\expandafter\ifx\csname natexlab\endcsname\relax\def\natexlab#1{#1}\fi

\bibitem[{{Alonso-Herrero} {et~al.}(2006){Alonso-Herrero}, {Perez-Gonzalez},
  {Alexander}, {Rieke}, {Rigopoulou}, {Le Floc'h}, {Barmby}, {Papovich},
  {Rigby}, {Bauer}, {Brandt}, {Egami}, {Willner}, {Dole}, \&
  {Huang}}]{alonso-herrero2006}
{Alonso-Herrero}, A., {Perez-Gonzalez}, P.~G., {Alexander}, D.~M., {et~al.}
  2006, \ApJ, 640, 167

\bibitem[{{Beelen} {et~al.}(2006){Beelen}, {Cox}, {Benford}, {Dowell},
  {Kovacs}, {Bertoldi}, {Omont}, \& {Carilli}}]{beelen2006}
{Beelen}, A., {Cox}, P., {Benford}, D.~J., {et~al.} 2006, \ApJ, 642, 694

\bibitem[{{Benford} {et~al.}(1999){Benford}, {Cox}, {Omont}, {Phillips}, \&
  {McMahon}}]{benford99}
{Benford}, D.~J., {Cox}, P., {Omont}, A., {Phillips}, T.~G., \& {McMahon},
  R.~G. 1999, \ApJ, 518, L65

\bibitem[{{Caputi} {et~al.}(2006){Caputi}, {Dole}, {Lagache}, {McLure},
  {Dunlop}, {Puget}, {Le Floc'h}, \& {P\'erez-Gonz\'alez}}]{caputi2006b}
{Caputi}, K.~I., {Dole}, H., {Lagache}, G., {et~al.} 2006, \AaA, 454, 143

\bibitem[{{Caputi} {et~al.}(2007){Caputi}, {Lagache}, {Yan}, {Dole},
  {Bavouzet}, {Le Floc'H}, {Choi}, {Helou}, \& {Reddy}}]{caputi2007}
{Caputi}, K.~I., {Lagache}, G., {Yan}, L., {et~al.} 2007, \ApJ, 660, 97

\bibitem[{{Chapman} {et~al.}(2005){Chapman}, {Blain}, {Smail}, \&
  {Ivison}}]{chapman2005}
{Chapman}, S.~C., {Blain}, A.~W., {Smail}, I., \& {Ivison}, R.~J. 2005, \ApJ,
  622, 772

\bibitem[{{Chapman} {et~al.}(2003){Chapman}, {Helou}, {Lewis}, \&
  {Dale}}]{chapman2003}
{Chapman}, S.~C., {Helou}, G., {Lewis}, G.~F., \& {Dale}, D.~A. 2003, \ApJ,
  588, 186

\bibitem[{{Chapman} {et~al.}(2004){Chapman}, {Smail}, {Blain}, \&
  {Ivison}}]{chapman2004}
{Chapman}, S.~C., {Smail}, I., {Blain}, A.~W., \& {Ivison}, R.~J. 2004, \ApJ,
  614, 671

\bibitem[{{Chary} \& {Elbaz}(2001)}]{chary2001}
{Chary}, R. \& {Elbaz}, D. 2001, \ApJ, 556, 562

\bibitem[{{Daddi} {et~al.}(2007){Daddi}, {Alexander}, {Dickinson}, {Gilli},
  {Renzini}, {Elbaz}, {Cimatti}, {Chary}, {Frayer}, {Bauer}, {Brandt},
  {Giavalisco}, {Grogin}, {Huynh}, {Kurk}, {Mignoli}, {Morrison}, {Pope}, \&
  {Ravindranath}}]{daddi2007a}
{Daddi}, E., {Alexander}, D.~M., {Dickinson}, M., {et~al.} 2007,
  Multiwavelength study of massive galaxies at z$\sim$2. II. Widespread Compton
  thick AGN and the concurrent growth of black holes and bulges,
  astro-ph/0705.2832

\bibitem[{{Dale} {et~al.}(2005){Dale}, {Bendo}, {Engelbracht}, {Gordon},
  {Regan}, {Armus}, {Cannon}, {Calzetti}, {Draine}, {Helou}, {Joseph},
  {Kennicutt}, {Li}, {Murphy}, {Roussel}, {Walter}, {Hanson}, {Hollenbach},
  {Jarrett}, {Kewley}, {Lamanna}, {Leitherer}, {Meyer}, {Rieke}, {Rieke},
  {Sheth}, {Smith}, \& {Thornley}}]{dale2005}
{Dale}, D.~A., {Bendo}, G.~J., {Engelbracht}, C.~W., {et~al.} 2005, \ApJ, 633,
  857

\bibitem[{{Dale} \& {Helou}(2002)}]{dale2002}
{Dale}, D.~A. \& {Helou}, G. 2002, \ApJ, 576, 159

\bibitem[{{Dole} {et~al.}(2006){Dole}, {Lagache}, {Puget}, {Caputi},
  {Fernández-Conde}, {Le Floc'h}, {Papovich}, {P\'erez-Gonz\'alez}, {Rieke},
  \& {Blaylock}}]{dole2006}
{Dole}, H., {Lagache}, G., {Puget}, J.~L., {et~al.} 2006, \AaA, 451, 417

\bibitem[{{Dole} {et~al.}(2004{\natexlab{a}}){Dole}, {Le Floc'h},
  {P\'erez-Gonz\'alez}, {Papovich}, {Egami}, {Lagache}, {Alonso-Herrero},
  {Engelbracht}, {Gordon}, {Hines}, {Krause}, {Misselt}, {Morrison}, {Rieke},
  {Rieke}, {Rigby}, {Young}, {Bai}, {Blaylock}, {Neugebauer}, {Beichman},
  {Frayer}, {Mould}, \& {Richards}}]{dole2004}
{Dole}, H., {Le Floc'h}, E., {P\'erez-Gonz\'alez}, P.~G., {et~al.}
  2004{\natexlab{a}}, \ApJS, 154, 87

\bibitem[{{Dole} {et~al.}(2004{\natexlab{b}}){Dole}, {Rieke}, {Lagache},
  {Puget}, {Alonso-Herrero}, {Bai}, {Blaylock}, {Egami}, {Engelbracht},
  {Gordon}, {Hines}, {Kelly}, {Le Floc'H}, {Misselt}, {Morrison}, {Muzerolle},
  {Papovich}, {Pérez-González}, {Rieke}, {Rigby}, {Neugebauer}, {Stansberry},
  {Su}, {Young}, {Beichman}, \& {Richards}}]{dole2004a}
{Dole}, H., {Rieke}, G.~H., {Lagache}, G., {et~al.} 2004{\natexlab{b}}, \ApJS,
  154, 93

\bibitem[{{Eisenhardt} {et~al.}(2004){Eisenhardt}, {Stern}, {Brodwin}, {Fazio},
  {Rieke}, {Rieke}, {Werner}, {Wright}, {Allen}, {Arendt}, {Ashby}, {Barmby},
  {Forrest}, {Hora}, {Huang}, {Huchra}, {Pahre}, {Pipher}, {Reach}, {Smith},
  {Stauffer}, {Wang}, {Willner}, {Brown}, {Dey}, {Jannuzi}, \&
  {Tiede}}]{eisenhardt2004}
{Eisenhardt}, P.~R., {Stern}, D., {Brodwin}, M., {et~al.} 2004, \ApJS, 154, 48

\bibitem[{{Fabricant} {et~al.}(2005){Fabricant}, {Fata}, {Roll}, {Hertz},
  {Caldwell}, {Gauron}, {Geary}, {McLeod}, {Szentgyorgyi}, {Zajac}, {Kurtz},
  {Barberis}, {Bergner}, {Brown}, {Conroy}, {Eng}, {Geller}, {Goddard},
  {Honsa}, {Mueller}, {Mink}, {Ordway}, {Tokarz}, {Woods}, {Wyatt}, {Epps}, \&
  {Dell'Antonio}}]{fabricant2005}
{Fabricant}, D., {Fata}, R., {Roll}, J., {et~al.} 2005, \PASP, 117, 1411

\bibitem[{{Fazio} {et~al.}(2004){Fazio}, {Ashby}, {Barmby}, {Hora}, {Huang},
  {Pahre}, {Wang}, {Willner}, {Arendt}, {Moseley}, {Brodwin}, {Eisenhardt},
  {Stern}, {Tollestrup}, \& {Wright}}]{fazio2004}
{Fazio}, G.~G., {Ashby}, M. L.~N., {Barmby}, P., {et~al.} 2004, \ApJS, 154, 39

\bibitem[{{Fiore} {et~al.}(2007){Fiore}, {Grazian}, {Santini}, {Puccetti},
  {Brusa}, {Feruglio}, {Fontana}, {Giallongo}, {Comastri}, {Gruppioni},
  {Pozzi}, {Zamorani}, \& {Vignali}}]{fiore2007}
{Fiore}, F., {Grazian}, A., {Santini}, P., {et~al.} 2007, Unveiling obscured
  accretion in the Chandra Deep Field South, astro-ph/0705.2864

\bibitem[{{Frayer} {et~al.}(2006){Frayer}, {Fadda}, {Yan}, {Marleau}, {Choi},
  {Helou}, {Soifer}, {Appleton}, {Armus}, {Beck}, {Dole}, {Engelbracht},
  {Fang}, {Gordon}, {Heinrichsen}, {Henderson}, {Hesselroth}, {Im}, {Kelly},
  {Lacy}, {Laine}, {Latter}, {Mahoney}, {Makovoz}, {Masci}, {Morrison},
  {Moshir}, {Noriega-Crespo}, {Padgett}, {Pesenson}, {Shupe}, {Squires},
  {Storrie-Lombardi}, {Surace}, {Teplitz}, \& {Wilson}}]{frayer2006}
{Frayer}, D.~T., {Fadda}, D., {Yan}, L., {et~al.} 2006, \AJ, 131, 250

\bibitem[{{Gordon}(2006)}]{gordon2006}
{Gordon}, K.~D. 2006, in prep.

\bibitem[{{Gordon} {et~al.}(2005){Gordon}, {Rieke}, {Engelbracht}, {Muzerolle},
  {Stansberry}, {Misselt}, {Morrison}, {Cadien}, {Young}, {Dole}, {Kelly},
  {Alonso-Herrero}, {Egami}, {Su}, {Papovich}, {Smith}, {Hines}, {Rieke},
  {Blaylock}, {Pérez-González}, {Le Floc'H}, {Hinz}, {Latter}, {Hesselroth},
  {Frayer}, {Noriega-Crespo}, {Masci}, {Padgett}, {Smylie}, \&
  {Haegel}}]{gordon2005}
{Gordon}, K.~D., {Rieke}, G.~H., {Engelbracht}, C.~W., {et~al.} 2005, \PASP,
  117, 503

\bibitem[{Jannuzi \& Dey(1999)}]{jannuzi99}
Jannuzi, B.~T. \& Dey, A. 1999, in ASP Conf. Ser. 191, Photometric Redshifts
  and High-Redshift Galaxies, ed. R.~Weymann, L.~Storrie-Lombardi, M.~Sawicki,
  \& R.~Brunner (San Francisco: ASP), 111

\bibitem[{{Kennicutt}(1998)}]{kennicutt98}
{Kennicutt}, R.~C. 1998, \ApJ, 498, 541

\bibitem[{{Kennicutt} {et~al.}(2003){Kennicutt}, {Armus}, {Bendo}, {Calzetti},
  {Dale}, {Draine}, {Engelbracht}, {Gordon}, {Grauer}, {Helou}, {Hollenbach},
  {Jarrett}, {Kewley}, {Leitherer}, {Li}, {Malhotra}, {Regan}, {Rieke},
  {Rieke}, {Roussel}, {Smith}, {Thornley}, \& {Walter}}]{kennicutt2003}
{Kennicutt}, R.~C., {Armus}, L., {Bendo}, G., {et~al.} 2003, \PASP, 115, 928

\bibitem[{{Kov\'acs} {et~al.}(2006){Kov\'acs}, {Chapman}, {Dowell}, {Blain},
  {Ivison}, {Smail}, \& {Phillips}}]{kovacs2006}
{Kov\'acs}, A., {Chapman}, S.~C., {Dowell}, C.~D., {et~al.} 2006, \ApJ, 650,
  592

\bibitem[{{Lacy} {et~al.}(2004){Lacy}, {Storrie-Lombardi}, {Sajina},
  {Appleton}, {Armus}, {Chapman}, {Choi}, {Fadda}, {Fang}, {Frayer},
  {Heinrichsen}, {Helou}, {Im}, {Marleau}, {Masci}, {Shupe}, {Soifer},
  {Surace}, {Teplitz}, {Wilson}, \& {Yan}}]{lacy2004}
{Lacy}, M., {Storrie-Lombardi}, L.~J., {Sajina}, A., {et~al.} 2004, \ApJS, 154,
  166

\bibitem[{{Lagache} {et~al.}(2004){Lagache}, {Dole}, {Puget},
  {P\'erez-Gonz\'alez}, {Le Floc'h}, {Rieke}, {Papovich}, {Egami},
  {Alonso-Herrero}, {Engelbracht}, {Gordon}, {Misselt}, \&
  {Morrison}}]{lagache2004}
{Lagache}, G., {Dole}, H., {Puget}, J.~L., {et~al.} 2004, \ApJS, 154, 112

\bibitem[{{Le F\`evre} {et~al.}(2005){Le F\`evre}, {Vettolani}, {Garilli},
  {Tresse}, {Bottini}, {Le Brun}, {Maccagni}, {Picat}, {Scaramella},
  {Scodeggio}, {Zanichelli}, {Adami}, {Arnaboldi}, {Arnouts}, {Bardelli},
  {Bolzonella}, {Cappi}, {Charlot}, {Ciliegi}, {Contini}, {Foucaud},
  {Franzetti}, {Gavignaud}, {Guzzo}, {Ilbert}, {Iovino}, {McCracken}, {Marano},
  {Marinoni}, {Mathez}, {Mazure}, {Meneux}, {Merighi}, {Paltani}, {Pellò},
  {Pollo}, {Pozzetti}, {Radovich}, {Zamorani}, {Zucca}, {Bondi}, {Bongiorno},
  {Busarello}, {Lamareille}, {Mellier}, {Merluzzi}, {Ripepi}, \&
  {Rizzo}}]{le_fevre2005}
{Le F\`evre}, O., {Vettolani}, G., {Garilli}, B., {et~al.} 2005, \AaA, 439, 845

\bibitem[{{Le Floc'h} {et~al.}(2005){Le Floc'h}, {Papovich}, {Dole}, {Bell},
  {Lagache}, {Rieke}, {Egami}, {P\'erez-Gonz\'alez}, {Alonso-Herrero}, {Rieke},
  {Blaylock}, {Engelbracht}, {Gordon}, {Hines}, {Misselt}, {Morrison}, \&
  {Mould}}]{le_floch2005}
{Le Floc'h}, E., {Papovich}, C., {Dole}, H., {et~al.} 2005, \ApJ, 632, 169

\bibitem[{{Le Floc'h} {et~al.}(2007){Le Floc'h}, {Willmer}, {Noeske},
  {Konidaris}, {Laird}, {Koo}, {Nandra}, {Bundy}, {Salim}, {Maiolino},
  {Conselice}, {Lotz}, {Papovich}, {Smith}, {Bai}, {Coil}, {Barmby}, {Ashby},
  {Huang}, {Blaylock}, {Rieke}, {Newman}, {Ivison}, {Chapman}, {Dole}, {Egami},
  \& {Elbaz}}]{le_floch2007}
{Le Floc'h}, E., {Willmer}, C. N.~A., {Noeske}, K., {et~al.} 2007, \ApJ, 660,
  L65

\bibitem[{{Marcillac} {et~al.}(2006){Marcillac}, {Elbaz}, {Chary}, {Dickinson},
  {Galliano}, \& {Morrison}}]{marcillac2006}
{Marcillac}, D., {Elbaz}, D., {Chary}, R.~R., {et~al.} 2006, \AaA, 451, 57

\bibitem[{{Papovich} {et~al.}(2006){Papovich}, {Cool}, {Eisenstein}, {Le
  Floc'h}, {Fan}, jr~{Kennicutt}, {Smith}, {Rieke}, \&
  {Vestergaard}}]{papovich2006}
{Papovich}, C., {Cool}, R., {Eisenstein}, D., {et~al.} 2006, \AJ, 132, 231

\bibitem[{{Papovich} {et~al.}(2004){Papovich}, {Dole}, {Egami}, {Le Floc'h},
  {P\'erez-Gonz\'alez}, {Alonso-Herrero}, {Bai}, {Beichman}, {Blaylock},
  {Engelbracht}, {Gordon}, {Hines}, {Misselt}, {Morrison}, {Mould},
  {Muzerolle}, {Neugebauer}, {Richards}, {Rieke}, {Rieke}, {Rigby}, {Su}, \&
  {Young}}]{papovich2004}
{Papovich}, C., {Dole}, H., {Egami}, E., {et~al.} 2004, \ApJS, 154, 70

\bibitem[{{Papovich} {et~al.}(2007){Papovich}, {Rudnick}, {Le Floc'H}, {van
  Dokkum}, {Rieke}, {Taylor}, {Armus}, {Gawiser}, {Huang}, {Marcillac}, \&
  {Franx}}]{papovich2007}
{Papovich}, C., {Rudnick}, G., {Le Floc'H}, E., {et~al.} 2007, \ApJ, 668, 45

\bibitem[{{Pope} {et~al.}(2006){Pope}, {Scott}, {Dickinson}, {Chary},
  {Morrison}, {Borys}, {Sajina}, {Alexander}, {Daddi}, {Frayer}, {Macdonald},
  \& {Stern}}]{pope2006}
{Pope}, A., {Scott}, D., {Dickinson}, M., {et~al.} 2006, \MNRAS, 370, 1185

\bibitem[{{Richards} {et~al.}(2006){Richards}, {Lacy}, {Storrie-Lombardi},
  {Hall}, {Gallagher}, {Hines}, {Fan}, {Papovich}, {Vanden Berk}, {Trammell},
  {Schneider}, {Vestergaard}, {York}, {Jester}, {Anderson}, {Budavári}, \&
  {Szalay}}]{richards2006}
{Richards}, G.~T., {Lacy}, M., {Storrie-Lombardi}, L.~J., {et~al.} 2006, \ApJS,
  166, 470

\bibitem[{{Rieke} {et~al.}(2004){Rieke}, {Young}, {Engelbracht}, {Kelly},
  {Low}, {Haller}, {Beeman}, {Gordon}, {Stansberry}, {Misselt}, {Cadien},
  {Morrison}, {Rivlis}, {Latter}, {Noriega-Crespo}, {Padgett}, {Stapelfeldt},
  {Hines}, {Egami}, {Muzerolle}, {Alonso-Herrero}, {Blaylock}, {Dole}, {Hinz},
  {Le Floc'H}, {Papovich}, {P\'erez-Gonz\'alez}, {Smith}, {Su}, {Bennett},
  {Frayer}, {Henderson}, {Lu}, {Masci}, {Pesenson}, {Rebull}, {Rho}, {Keene},
  {Stolovy}, {Wachter}, {Wheaton}, {Werner}, \& {Richards}}]{rieke2004}
{Rieke}, G.~H., {Young}, E.~T., {Engelbracht}, C.~W., {et~al.} 2004, \ApJS,
  154, 25

\bibitem[{{Sajina} {et~al.}(2006){Sajina}, {Scott}, {Dennefeld}, {Dole},
  {Lacy}, \& {Lagache}}]{sajina2006}
{Sajina}, A., {Scott}, D., {Dennefeld}, M., {et~al.} 2006, \MNRAS, 369, 939

\bibitem[{{Sanders} \& {Mirabel}(1996)}]{sanders96}
{Sanders}, D.~B. \& {Mirabel}, I.~F. 1996, \ARAA, 34, 749

\bibitem[{{Soifer} {et~al.}(1987){Soifer}, {Sanders}, {Madore}, {Neugebauer},
  {Danielson}, {Elias}, {Lonsdale}, \& {Rice}}]{soifer87}
{Soifer}, B.~T., {Sanders}, D.~B., {Madore}, B.~F., {et~al.} 1987, \ApJ, 320,
  238

\bibitem[{{Stern} {et~al.}(2005){Stern}, {Eisenhardt}, {Gorjian}, {Kochanek},
  {Caldwell}, {Eisenstein}, {Brodwin}, {Brown}, {Cool}, {Dey}, {Green},
  {Jannuzi}, {Murray}, {Pahre}, \& {Willner}}]{stern2005}
{Stern}, D., {Eisenhardt}, P., {Gorjian}, V., {et~al.} 2005, \ApJ, 631, 163

\bibitem[{{Swinbank} {et~al.}(2004){Swinbank}, {Smail}, {Chapman}, {Blain},
  {Ivison}, \& {Keel}}]{swinbank2004}
{Swinbank}, A.~M., {Smail}, I., {Chapman}, S.~C., {et~al.} 2004, \ApJ, 617, 64

\bibitem[{{Symeonidis} {et~al.}(2006){Symeonidis}, {Rigopoulou}, {Huang}, \&
  {Davis}}]{symeonidis2006}
{Symeonidis}, M., {Rigopoulou}, D., {Huang}, J.-S., \& {Davis}, M. 2006, \ApJ

\bibitem[{{Takeuchi} {et~al.}(2005){Takeuchi}, {Buat}, {Iglesias-P\'aramo},
  {Boselli}, \& {Burgarella}}]{takeuchi2005a}
{Takeuchi}, T.~T., {Buat}, V., {Iglesias-P\'aramo}, J., {Boselli}, A., \&
  {Burgarella}, D. 2005, \AaA, 432, 423

\bibitem[{{Taylor} {et~al.}(2005){Taylor}, {Mann}, {Efstathiou}, {Babbedge},
  {Rowan-Robinson}, {Lagache}, {Lawrence}, {Mei}, {Vaccari}, {Héraudeau},
  {Oliver}, {Dennefeld}, {Perez-Fournon}, {Serjeant}, {González-Solares},
  {Puget}, {Dole}, \& {Lari}}]{taylor2005}
{Taylor}, E.~L., {Mann}, R.~G., {Efstathiou}, A.~N., {et~al.} 2005, \MNRAS,
  361, 1352

\bibitem[{{Vanzella} {et~al.}(2005){Vanzella}, {Cristiani}, {Dickinson},
  {Kuntschner}, {Moustakas}, {Nonino}, {Rosati}, {Stern}, {Cesarsky}, {Ettori},
  {Ferguson}, {Fosbury}, {Giavalisco}, {Haase}, {Renzini}, {Rettura}, {Serra},
  \& {The Goods Team}}]{vanzella2005}
{Vanzella}, E., {Cristiani}, S., {Dickinson}, M., {et~al.} 2005, \AaA, 434, 53

\bibitem[{{Vanzella} {et~al.}(2006){Vanzella}, {Cristiani}, {Dickinson},
  {Kuntschner}, {Nonino}, {Rettura}, {Rosati}, {Vernet}, {Cesarsky},
  {Ferguson}, {Fosbury}, {Giavalisco}, {Grazian}, {Haase}, {Moustakas},
  {Popesso}, {Renzini}, {Stern}, \& {The Goods Team}}]{vanzella2006}
{Vanzella}, E., {Cristiani}, S., {Dickinson}, M., {et~al.} 2006, \AaA, 454, 423

\bibitem[{{Werner} {et~al.}(2004){Werner}, {Roellig}, {Low}, {Rieke}, {Rieke},
  {Hoffmann}, {Young}, {Houck}, {Brandl}, {Fazio}, {Hora}, {Gehrz}, {Helou},
  {Soifer}, {Stauffer}, {Keene}, {Eisenhardt}, {Gallagher}, {Gautier}, {Irace},
  {Lawrence}, {Simmons}, {van Cleve}, {Jura}, {Wright}, \&
  {Cruikshank}}]{werner2004}
{Werner}, M.~W., {Roellig}, T.~L., {Low}, F.~J., {et~al.} 2004, \ApJS, 154, 1

\bibitem[{{Wolf} {et~al.}(2004){Wolf}, {Meisenheimer}, {Kleinheinrich},
  {Borch}, {Dye}, {Gray}, {Wisotzki}, {Bell}, {Rix}, {Cimatti}, {Hasinger}, \&
  {Szokoly}}]{wolf2004}
{Wolf}, C., {Meisenheimer}, K., {Kleinheinrich}, M., {et~al.} 2004, \AaA, 421,
  913

\bibitem[{{Yang} {et~al.}(2007){Yang}, {Greve}, {Dowell}, \&
  {Borys}}]{yang2007}
{Yang}, M., {Greve}, T.~R., {Dowell}, C.~D., \& {Borys}, C. 2007, \ApJ, 660,
  1198

\bibitem[{{Yang} \& {Phillips}(2007)}]{yang2007a}
{Yang}, M. \& {Phillips}, T. 2007, \ApJ, 662, 284

\bibitem[{{Zheng} {et~al.}(2007){Zheng}, {Dole}, {Bell}, {Le Floc'H}, {Rieke},
  {Rix}, \& {Schiminovich}}]{zheng2007}
{Zheng}, X.~Z., {Dole}, H., {Bell}, E.~F., {et~al.} 2007, Infrared Spectral
  Energy Distributions of z$\sim$0.7 Star-Forming Galaxies, astro-ph/0706.0003

\end{thebibliography}


\begin{appendix}


\section{The K-corrections for the stacking points}
\label{app:kcorr_stacking}

\begin{table*}[t!]
\caption{Maximal errors obtained with the \citet{dale2002} templates
when doing a K-correction for redshift averaged bins instead of a 
K-correction weighted by the redshift distribution.}
\label{table:average_kcorr}
\centering
\begin{tabular}{c c c c c c c}
\hline\hline
$S_{24\mu\textrm{m}}$ bin (mJy) & Redshift bin & 8~$\mu$m & 24~$\mu$m & 70~$\mu$m & 160~$\mu$m \\
\hline
                 & $0<z<0.25$   &  5.4\% & 1.2\% & 0.7\% & 1.3\% \\
$0.8<S_{24}<1.5$ & $0.25<z<0.5$ &  3.5\% & 1.5\% & 1.1\% & 1.3\% \\
                 & $0.5<z<1$    &  8.0\% & 1.5\% & 0.8\% & 0.6\% \\
\hline
                 & $0<z<0.25$   &  4.8\% & 0.8\% & 0.6\% & 1.0\% \\
 $1.5<S_{24}<3$  & $0.25<z<0.5$ &  4.0\% & 1.0\% & 0.9\% & 1.0\% \\
                 & $0.5<z<1$    & 10.6\% & 3.1\% & 2.3\% & 1.5\% \\
\hline
                 & $0<z<0.25$   &  5.1\% & 1.0\% & 0.9\% & 1.2\% \\
  $3<S_{24}<10$  & $0.25<z<0.5$ &  2.8\% & 0.8\% & 0.7\% & 0.5\% \\
                 & $0.5<z<1$    & $\cdots$ & $\cdots$ & $\cdots$ & $\cdots$ \\
\hline
\end{tabular}
\end{table*}

In Sect.~\ref{section:kcorr} we K-corrected the stacking points using
the mean redshift of the sources rather than averaging
over the redshift distribution of the bin. We used simulations to
test whether this simplification significantly affects our results. 

For each template in the \citet{dale2002} library we built a mock
sample with the redshift and $S_{24}$ flux distributions of our
Bo\"otes sub-samples (see Tab.~\ref{table:stack_bootes}), computing
both the rest-frame and observed-frame 8, 24, 70 and 160~$\mu$m fluxes
for each source. We then compared the true, averaged rest-frame 
fluxes to the values found by averaging the observed-frame fluxes
and K-correcting to the rest frame using the average redshift as
we do in the stacking analysis. The results are
given in Tab.~\ref{table:average_kcorr}. The differences
are negligible at  24, 70 and 160~$\mu$m. At 8~$\mu$m, the K-corrections 
vary strongly with redshift because of the PAH features. As a result,
the errors are larger and reach $\sim$10\% for the two $0.5<z<1$ bins. 
However, these two points have such large uncertainties in our data
(see Tab.~\ref{table:stack_bootes}), that even 10\% correction would not
significantly change our results. We conclude that using the 
K-correction corresponding to the mean bin redshift is sufficiently
accurate for our purposes.


\section{Effect of the selection on the $\nu L_\nu - L_\textrm{IR}$ correlations}
\label{annex}

We used a simulation to understand and quantify
the effects of selection criteria on the 
correlations presented in this paper.

\begin{figure}[t!]
\resizebox{\hsize}{!}{\includegraphics{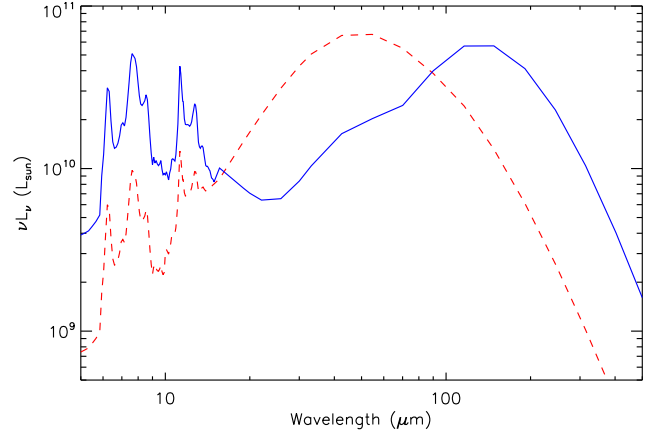}}
\caption{Spectral energy distribution of the two templates from the
\citet{dale2002} library used in the
simulation. The solid blue line corresponds to $\alpha = 3.5$ (the cold template)
and the dashed red line corresponds to $\alpha = 1.3$ (the warm template). Both
templates are normalized to the same total infrared luminosity 
($L_\textrm{IR}=10^{11} L_\odot$).}
\label{figure:simu_templates}
\end{figure}

We used two templates characterized by their $\alpha$ parameter
from the \citet{dale2002} library to model the galaxies.
\citet{dale2005} found that most galaxies have SEDs in 
the range $1.3 < \alpha < 3.5$ so we use the two
extremes of a {\it warm} template with $\alpha = 1.3$
and a {\it cold} template with $\alpha = 3.5$. 
These two templates are shown in Fig.~\ref{figure:simu_templates}.

We randomly and uniformly distributed 20000 galaxies in the
$z-\log(L_\textrm{IR})$ plane over the range $0<z<1$, 
$10 < \log(\frac{L_\textrm{IR}}{L_\odot}) < 13$, and
$L_\textrm{IR} < 10^{13} \times z^{1.42}$. The upper
limit on the luminosity was determined empirically from
our sample of 372 directly detected galaxies. We randomly 
assigned half of the galaxies to be warm and the other
half to be cold and then computed the 8, 24, 70 and 160~$\mu$m
fluxes of each galaxy. Given a set of detection limits,
we can now explore which kind of galaxies will be detectable.

\begin{figure}[t!]
\resizebox{\hsize}{!}{\includegraphics{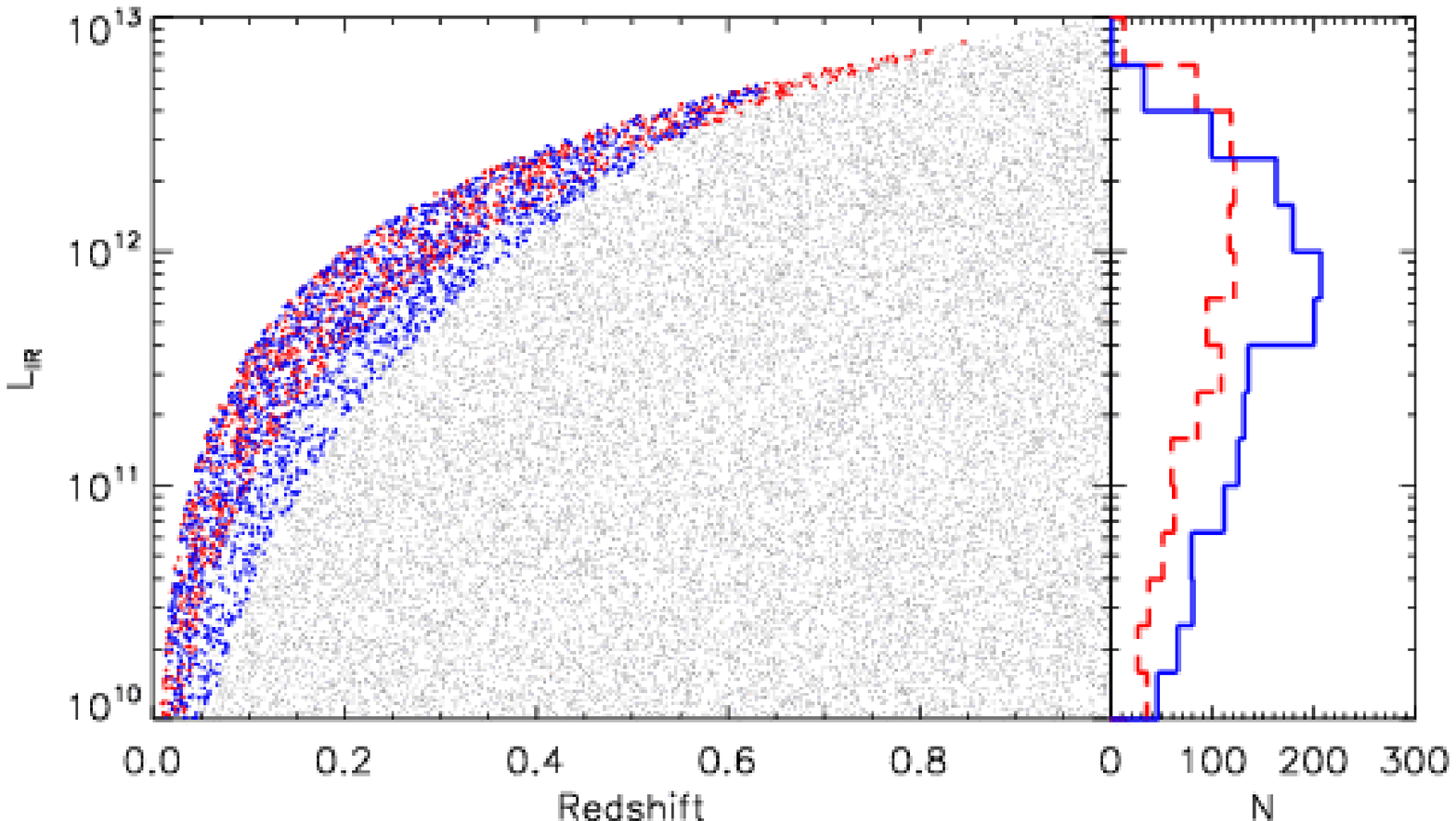}}
\resizebox{\hsize}{!}{\includegraphics{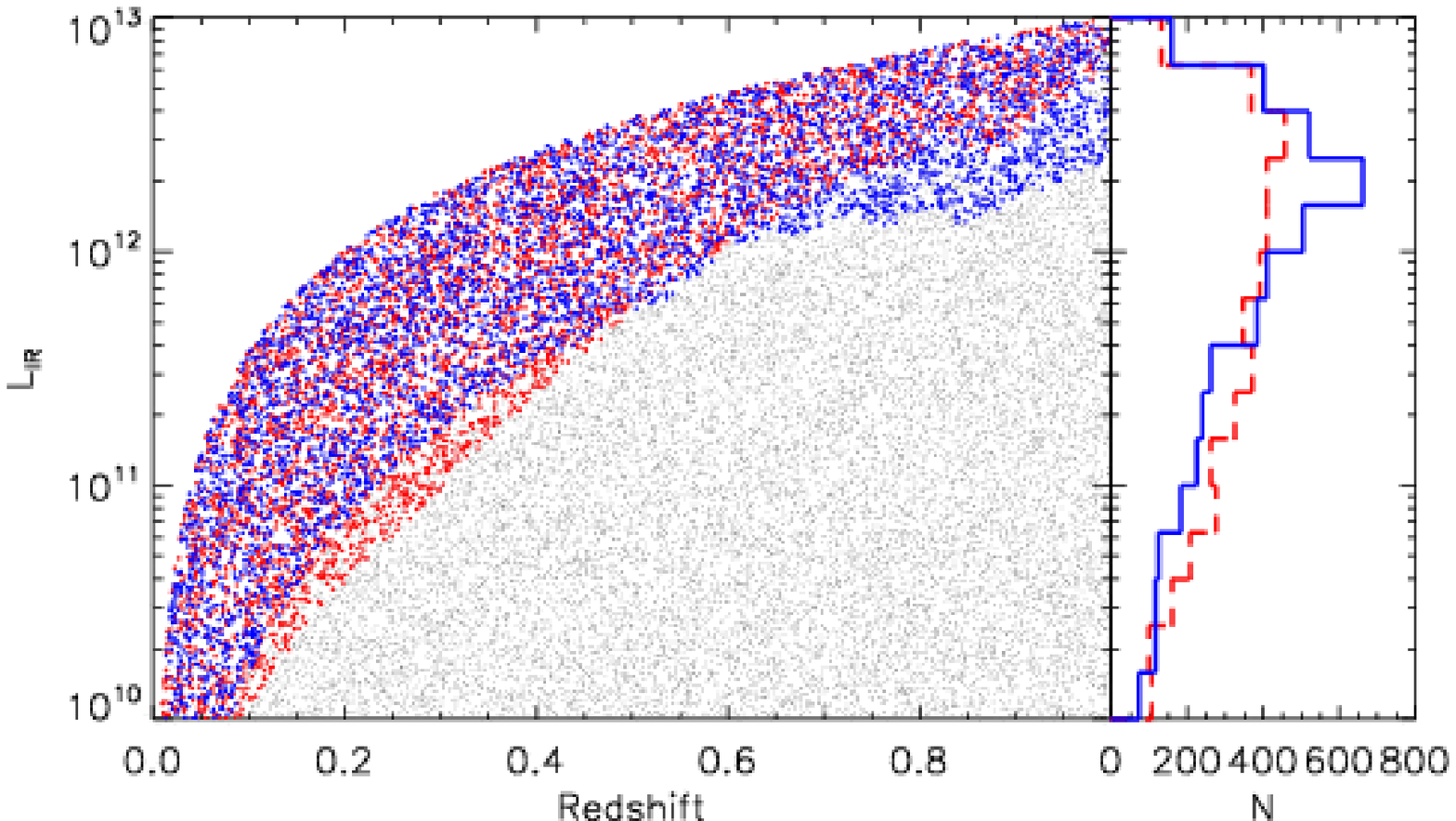}}
\caption{Results of the simulation with two different sets of detection limits.
The black dots correspond to the 20000 galaxies in our simulation (both warm and cold).
The red and blue circles indicate warm and cold detected galaxies, respectively. 
{\it Upper panel:} Direct detections using limits of [0.006, 1, 23, 92] (mJy) at
8, 24, 70 and 160~$\mu$m, respectively. {\it Lower panel:} Stacking detections using
the limits of [0.006, 1, 2.3, 9.2] (mJy). In both panels, the histograms show the
distributions of detected warm (red) and cold (blue) galaxies as a function of 
infrared luminosity.}
\label{figure:simu_selection}
\end{figure}

In the Bo\"otes field, the difference is between the 
directly detected sources and the stacked sources. The 
directly detected sources had to exceed flux limits of
[0.006, 1, 23, 92] (mJy) at 8, 24, 70 and 160~$\mu$m, respectively.
The stacked sources had to be detected at 24~$\mu$m (and effectively
at 8~$\mu$m) but for a stack of 100 sources they could be 10 times
fainter in the longer wavelength bands. Thus, the stacked sources
had to exceed flux limits of [0.006, 1, 2.3, 9.2] (mJy).
Figure~\ref{figure:simu_selection} illustrates the consequences
of these two selection criteria on the balance between the 
warm and cold sources as a function of luminosity.
For normal galaxies with $L_\textrm{IR} < 10^{11} L_\odot$, directly
detected galaxies tend to be cold while stacked galaxies tend to be
warm. For high luminosities ($L_\textrm{IR} > 10^{12} L_\odot$),
we see the reverse. We probe colder sources with the stacking
analysis than with direct detection. Both effects are a 
consequence of the warm galaxies having a higher $(\nu L_\nu)_{24\mu \textrm{m, rest}}$ 
and lower $(\nu L_\nu)_{160\mu \textrm{m, rest}}$ than the cold galaxies
(see Fig.~\ref{figure:simu_templates}). Thus, by stacking the undetected 
far-infrared galaxies, we are preferentially adding
warm galaxies at low luminosities and cold galaxies at high luminosities,
and this explains the differences observed in Fig.~\ref{figure:correl_stacking}.

\section{Influence of the library choice on the K-corrections}
\label{app:kcorr_model}

\begin{figure}
\resizebox{\hsize}{!}{\includegraphics{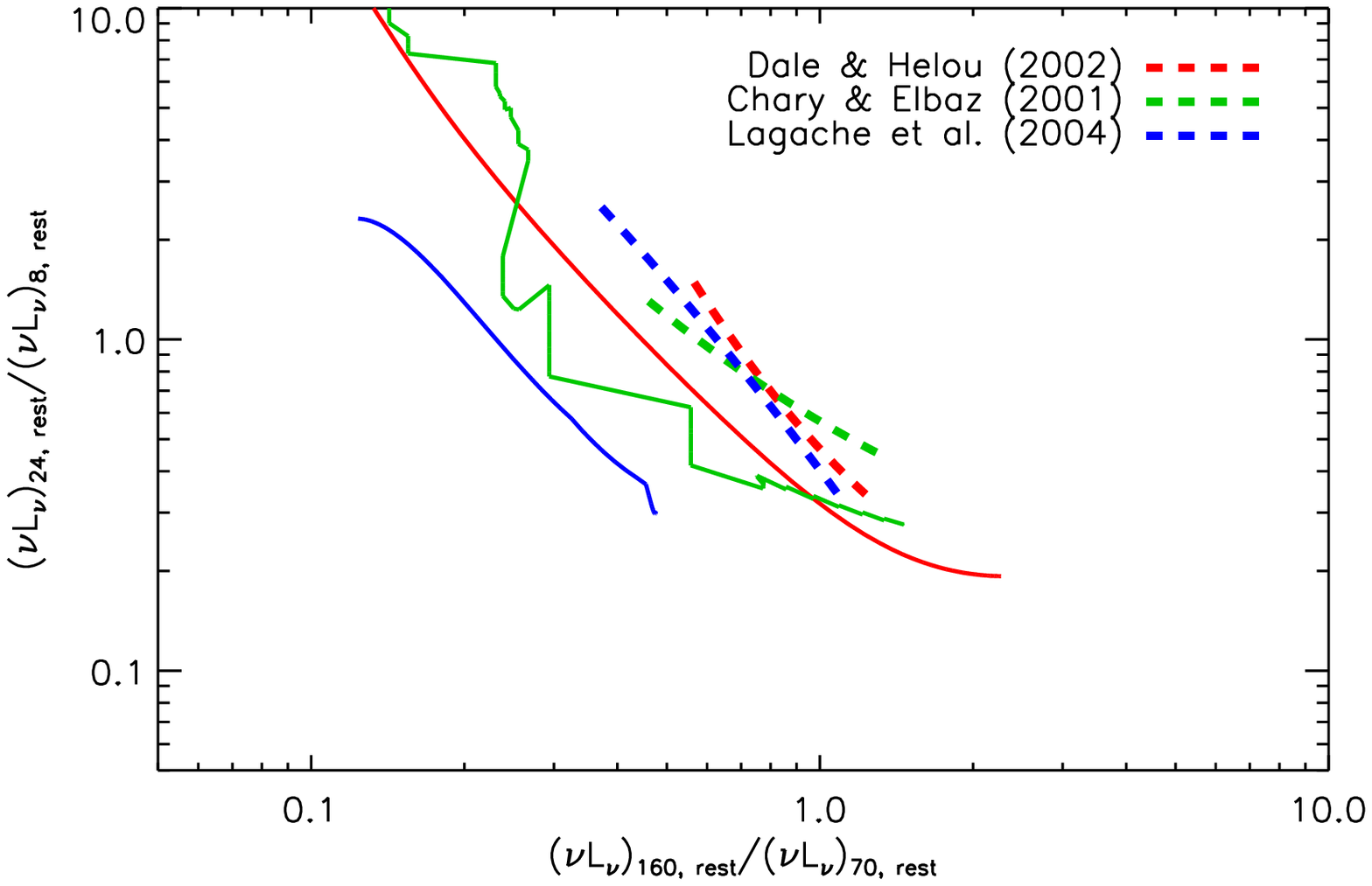}}
\caption{As in Fig.~\ref{figure:evol_color}, but showing the
median restframe colors (thick dashed lines) of the directly detected galaxies after
computing the K-corrections with three different templates 
\citep{dale2002,chary2001,lagache2004}.}
\label{fig:kcorr_model}
\end{figure}

While the infrared colors of our sample are in better agreement with the predictions
of the \citet{dale2002} model (Fig.~\ref{figure:evol_color}), one might argue
that this is a consequence of using these templates to compute
the K-corrections. However, we have verified that using the 
\citet{lagache2004,chary2001} libraries for the K-corrections has no effect 
on our conclusion. Figure~\ref{fig:kcorr_model} shows the median rest-frame 
colors computed with three different SED models \citep{dale2002,chary2001,lagache2004} 
for the 372 galaxies directly detected at all wavelengths. We clearly see that
for all three K-correction models the resulting rest-frame colors of the 
sample are in better agreement with the \citet{dale2002} model predictions.

\end{appendix}

\end{document}